\documentclass[twocolumn]{aastex61}

\newcommand{\lya}{Ly$\alpha$}

\newcommand{\ha}{H$\alpha$}

\newcommand{\hb}{H$\beta$}

\newcommand{\hi}{H{\sc i}}
\newcommand{\hii}{H{\sc ii}}
\newcommand{\nii}{[N{\sc ii}]}
\newcommand{\cii}{C\,{\sc ii}}
\newcommand{\ciii}{C\,{\sc iii}}
\newcommand{\sii}{[S{\sc ii}]}
\newcommand{\siii}{[S{\sc iii}]}
\newcommand{\oi}{O\,{\sc i}}
\newcommand{\oii}{[O{\sc ii}]}
\newcommand{\oiii}{[O{\sc iii}]}
\newcommand{\sit}{Si\,{\sc ii}}
\newcommand{\sif}{Si\,{\sc iv}}
\newcommand{\oratio}{[O\,{\sc iii}]/[O\,{\sc ii}]}

\newcommand{\egs}{erg/s/cm$^{2}$}
\newcommand{\egsa}{erg/s/cm$^{2}$/\AA}

\newcommand{\msun}{$M_\odot$}
\newcommand{\msunyr}{$M_\odot$/year}
\newcommand{\fesclyc}{$f_\mathrm{esc}^{\mathrm{LyC}}$}
\newcommand{\fesclya}{$f_\mathrm{esc}^{\mathrm{Ly}\alpha}$}

\newcommand{\haro}{Haro\,11}

\received{Sep 11, 2020}
\revised{Mar 9, 2021}
\accepted{Mar 23, 2021}
\submitjournal{ApJ}

\shorttitle{The source of leaking ionizing photons  from \haro }
\shortauthors{\"Ostlin, Rivera-Thorsen, Menacho, Hayes et al.}

\begin{document}

\title{The source of leaking ionizing photons  from Haro11 -- Clues from HST/COS spectroscopy of knots A, B and C \footnote{Based of observations with the NASA/ESA Hubble Space Telescope.}}


\correspondingauthor{G\"oran \"Ostlin}
\email{ostlin@astro.su.se}

\author[0000-0002-3005-1349]{G\"oran \"Ostlin} \affil{Department of Astronomy, Oskar Klein Centre; Stockholm University; 106\,91 Stockholm, Sweden}
\author{T. Emil Rivera-Thorsen} \affil{Department of Astronomy, Oskar Klein Centre; Stockholm University; 106\,91 Stockholm, Sweden}
\author{Veronica Menacho} \affil{Department of Astronomy, Oskar Klein Centre; Stockholm University; 106\,91 Stockholm, Sweden}
\author{Matthew Hayes} \affil{Department of Astronomy, Oskar Klein Centre; Stockholm University; 106\,91 Stockholm, Sweden}
\author{Axel Runnholm} \affil{Department of Astronomy, Oskar Klein Centre; Stockholm University; 106\,91 Stockholm, Sweden}
\author{Genoveva Micheva} \affil{Leibniz-Institute for Astrophysics Potsdam, An der Sternwarte 16, 14482 Potsdam, Germany}
\author{M. S. Oey} \affil{University of Michigan, Department of Astronomy, 323 West Hall, 1085 South University Ave, Ann Arbor, MI 48109-1107, USA}
\author{Angela Adamo} \affil{Department of Astronomy, Oskar Klein Centre; Stockholm University; 106\,91 Stockholm, Sweden}
\author{Arjan Bik} \affil{Department of Astronomy, Oskar Klein Centre; Stockholm University; 106\,91 Stockholm, Sweden}
\author{John M. Cannon} \affil{Department of Physics \& Astronomy, Macalester College, 1600 Grand Avenue, Saint Paul, MN 55105, USA}
\author{Max Gronke}\thanks{Hubble fellow} \affil{Department of Physics \& Astronomy, Johns Hopkins University, 
    Bloomberg Center, 3400 N. Charles St., Baltimore, MD 21218, USA}
\author{Daniel Kunth} \affil{Institut d'Astrophysique de Paris, 98bis Boulevard Arago, 75014 Paris, France}
\author{Peter Laursen} \affil{Cosmic Dawn Center (DAWN)}
\affil{ Niels Bohr Institute, University of Copenhagen, Jagtvej 128, DK-2200, Copenhagen N, Denmark}
\author{Miguel Mas-Hesse} \affil{Centro de Astrobiolog\'{\i}a (CSIC--INTA), Depto. de Astrof\'{\i}sica, 28692 Villanueva de la Ca\~nada, Madrid, Spain}
\author{Jens Melinder} \affil{Department of Astronomy, Oskar Klein Centre; Stockholm University; 106\,91 Stockholm, Sweden}
\author{Matteo Messa} \affil{University of Massachusetts, Department of Astronomy, 710 North Pleasant Street Amherst, MA 01003-9305, USA}
\author{Mattia Sirressi} \affil{Department of Astronomy, Oskar Klein Centre; Stockholm University; 106\,91 Stockholm, Sweden}
\author{Linda Smith} \affil{European Space Agency (ESA), ESA Office, Space Telescope Science Institute, 3700 San Martin Drive, Baltimore, MD 21218, USA}

      




\begin{abstract}

Understanding the escape of ionizing (Lyman continuum) photons  from galaxies is vital for 
determining how galaxies contributed to reionization in the early universe. 
While directly detecting  Lyman continuum from high redshift galaxies  is impossible due to the intergalactic medium,
 low redshift galaxies in principle offer this possibility, but  requirie observations from space. 
 The first local galaxy for which Lyman continuum escape was found is \haro , a luminous blue compact 
 galaxy at  $z=0.02$,  where observations with the  FUSE satellite revealed an  escape fraction of 3.3\%.
 However the FUSE aperture covers the entire
 galaxy, and it is not clear from where  the  Lyman continuum is leaking out. 
 Here we utilize HST/COS spectroscopy in the wavelength range 1100-1700 \AA\ of the three  knots 
 (A, B, and C) of \haro\ to study the presence of \lya\ emission and the properties of intervening  gas. 
 We find that all knots have bright \lya\ emission. UV absorption lines, originating in the neutral interstellar 
 medium, as well as  lines probing the ionized medium, are seen extending to blue shifted velocities
 of 500 km/s  in all three knots, demonstrating the presence of an outflowing multiphase medium.
We find that knots A and B have  large covering fractions of neutral gas, making LyC escape 
 along these sightlines  improbable, while  knot C has a much lower covering 
 fraction ($\lesssim50$\%). 
 Knot C also has the  the highest \lya\ escape fraction and we conclude that it is
the most  likely source of the escaping Lyman continuum  detected in \haro .     

\end{abstract}
\keywords{cosmology: observations --- galaxies: star-burst --- galaxies: individual (\haro )}

\section{Introduction}
\label{intro}

Understanding how galaxies  have contributed to the reionization of the universe is a major theme in contemporary astrophysics. 
In the quest for this, it is vital to know under what conditions ionizing (Lyman continuum) radiation can make their way through the interstellar
medium (ISM) and escape from galaxies.  Directly observing Lyman continuum (LyC) escape in the epoch of reionization (EoR, $z>6$) and even at 
somewhat lower redshifts ($z>4$) is impossible due to the long path lengths and increasing probability of intersecting neutral clouds in  
the intergalactic medium \citep[IGM,][]{inoue2008}.
Photometric surveys have been used to find candidate LyC leakers at $z\sim3$ \citep[e.g.][]{iwata2009,mostardi2013,micheva2017}, but such surveys are prone to 
contamination by lower redshift sources and followup studies are typically able to confirm only very few candidates selected by photometry
as real LyC leakers \citep{vanzella2012,siana2015,mostardi2015}.
At lower redshifts ($z\lesssim2$) IGM absorption is less of an obstacle, but with the exception of rare objects strongly  magnified by gravitational
lensing \citep[e.g. the {\it Sunburst arc} at $z=2.4$,][]{rivera-thorsen2019}, sources are still largely unresolved, preventing a detailed study of the escape mechanism. Galaxies in the 
local universe, on the other hand,  offer the possibility of  detailed studies   of processes that could lead to Lyman continuum escape, but
here the challenge is to actually measure the escape of photons with a rest frame wavelength in the  far ultraviolet, $\lambda\sim 900$\AA .

Sensitivity to such wavelengths was provided with the FUSE satellite, and the first local galaxy for which a net escape of Lyman continuum
radiation was claimed was \haro\    \citep{bergvall2006}. This claim was disputed by  \citet{grimes2007}, but a thorough reanalysis confirmed the  
leakage \citep{leitet2011} and an escape fraction of 3.3\% was derived. 
However, the large ($30\arcsec\times30\arcsec$)  aperture of the FUSE spectrograph covers the entire galaxy, leaving the question "From where 
in the galaxy is Lyman continuum leaking out?" unanswered. 
HST/COS \citep{green2012} is also capable of measuring spectra down to the rest frame Lyman limit \citep{mccandliss2010}, but has very poor sensitivity below 1100\AA , making 
observations quite costly in terms of exposure time, but some leaking galaxies have nevertheless been confirmed \citep{leitherer2016, puschnig2017}. 
Subsequently, HST/COS has been successfully explored to measure LyC escape from compact galaxies at $z\gtrsim 0.2$ \citep{izotov2018a, izotov2018b} 
where the COS sensitivity is  much higher. These sources, mostly of the green pea galaxy type \citep{cardamone2009} selected to have high 
ionization (emission line ratios \oratio $>5$) are though 
barely resolved, preventing an analysis of the details of the escape mechanism. 

\haro\ is a luminous blue compact galaxy (BCG) at $z=0.0206$ with a starburst triggered by a dwarf galaxy merger \citep{ostlin2001, ostlin2015}. 
It features an irregular morphology with three bright knots: A, B and C \citep[see Fig. \ref{fig:fig1} and][]{vader1993}. It has a low metallicity \citep[oxygen abundance $\sim 20\%$ solar,][]{bo2002}
 but with    indications that the oxygen abundance differs significantly between the knots by more than 0.2 dex 
 \citep{guseva2012, james2013, menacho2021}. Its star formation 
rate is in the range 20 to 30 \msunyr\ depending on the tracer used  \citep{ostlin2015}. \haro\ is remarkably devoid of neutral hydrogen, with only
$5\times 10^8$\msun\ \citep{pardy2016,machattie2014} detected, and the molecular mass (albeit quite uncertain) is of the order
  $\gtrsim 10^9$\msun\ \citep{cormier2014}. 
The gas consumption time scale is hence on the order of 50 Myr, making \haro\ one of the most extreme starbursts known. Bright X-ray point sources 
are associated with knots B and C \citep{grimes2007, prestwich2015, basu-zych2016, dittenber2020}. \lya\ imaging of \haro\ showed 
bright emission from knot C, with extensions to north and south indicating a possible outflow origin, and large scale diffuse emission on a galaxy 
wide scale, but nothing from knots A and B \citep{hayes2007,ostlin2009}. Since  the escape of \lya , just like that of LyC, is  sensitive to the amount of neutral hydrogen,
\citet{hayes2007}  suggested that the escaping LyC radiation was likely to come from knot C.  
Being the nearest known LyC  emitter, \haro\ offers
unique opportunities to study  the details of LyC escape and how indirect LyC diagnostics behave in a spatially resolved fashion.
  
Knot C was observed with HST/COS in gratings G130M and G160M \citep{alexandroff2015,heckman2015} shoving evidence of outflows of 
neutral gas probed by low ionization state (LIS) metal lines, e.g. from \sit .  These spectra were 
reanalyzed by \citet{rivera-thorsen2017} using the apparent optical depth (AOD) method \citep{ss1991, jones2013, 
rivera-thorsen2015} applied to     \sit\ lines   of different oscillator strengths to solve for both the column density and 
covering fraction of the absorbing material, as a function of velocity. Since silicon has a lower ionization energy (8.15 eV) than hydrogen (16.6 eV), \sit\ traces 
gas where hydrogen is mostly neutral, while \sif\ traces highly ionized gas (comparable to \oiii ).  These results showed that knot C has a 
low covering ($\sim 50\%$) of neutral gas clouds over a wide velocity range, along the sight line corraborating it as a good candidate for the
source of ionizing photons observed with FUSE. 

In the classical picture, an \hii\ region that is radiation bounded 
 will exhibit a highly ionized inner \oiii\ zone, and an outer \oii\ zone (where rarer photons capable of doubly ionizing oxygen have been
exhausted) \citep{agn2}. If, on the other hand, there are enough photons to completely ionize the local ISM, the \hii\ region is said to be density bounded,
and  ionizing photons will leak out into the galactic scale ISM. In this case, the ionization state of the gas, as traced by \oratio\  will remain high
\citep{pellegrini2012,jaskot2013}.
This is a factor for the successful COS observations at $z\gtrsim0.2$ of Lyman continuum escape from galaxies selected to have a high global 
\oratio\ ratios from SDSS \citep{izotov2018a, izotov2018b}.
\haro\ was imaged by the HST in the \oii\ and \oiii\ lines, with the scope to use this diagnostic to pinpoint likely sites of the escaping Lyman
continuum radiation. With knot C showing low ionization, knot B moderate, and knot A  large scale high ionization, 
\citet{keenan2017}  proposed  knot A as the most probable  source of leaking Lyman continuum.
 
Here we present  new HST/COS observations with gratings G130M and G160M of knots A and B. We characterize the \lya\ emission and 
the \sit\ and \sif\ transitions, along with knot C,  and include optical emission line diagnostics from ESO/VLT/MUSE to provide further insight 
into the question on which source(s) that may dominate the observed Lyman continuum escape

Throughout the paper we assume  a luminosity 
distance of 88 Mpc ($m-M=34.7$) and a scale of 0.42 kpc/arcsec.

\section{Observations and reductions}
\subsection{COS data}
\begin{figure}  
\centering
\includegraphics[angle=0,scale=0.41,trim=0cm 5cm 0 5cm]{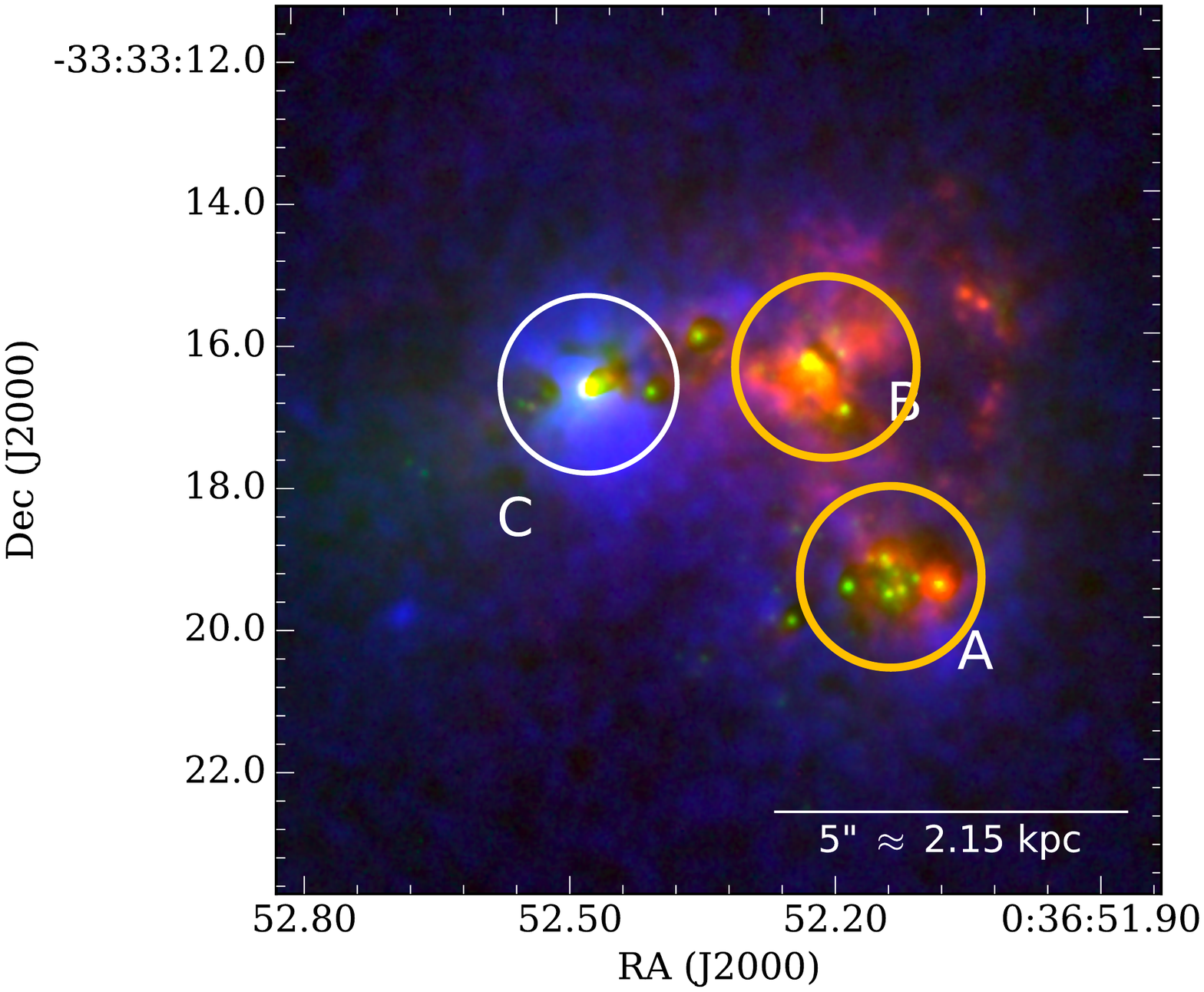}
\caption{RGB composite of \haro\ with \lya\ in blue, UV-continuum ($\lambda\sim1500$ \AA) in green, and \ha\ in red \citep{ostlin2009, rivera-thorsen2017}. 
The white circle show the COS aperture for the previously published spectrum of knot C  \citep{alexandroff2015,rivera-thorsen2017}, 
and the orange circles the COS apertures for the new spectra presented in this paper.} 
\label{fig:fig1}
\end{figure}

We have used archival  HST/COS observations in G130M and G160M of knot C from program 13017 (PI Heckman), and new COS 
observations of knots A and B with the same gratings (from program 15352, PI Oestlin). 
Knot C was used for acquisition, using 'mirror A' \citep[see COS instrument handbook, IHB, ][]{cos-ihb}, followed by blind offsets to 
knots A and B. The reason for this was that knot C is  a bright and relatively isolated
 point-like source in the UV , while knots A and B have a more complicated structure encompassing several star clusters. 
 Hence acquiring on knot C, followed by blind offsets,  was judged to produce a more predictable location of the  apertures on knot A and B. 
 Each of knots A and B were observed in single visits of 2 orbit durations.
Acquisition images  of 51, and 60 seconds were obtained for knots A and B, respectively.
The new observations use COS lifetime position 4 \citep[see COS IHB,][]{cos-ihb}. Four different wavelength/FP-POS settings
per observation were used to cover detector gaps, optimized to get the best coverage of the features of interest. In order to protect
the COS detector (e.g. from geocoronal \lya ), not all possible  combinations were allowed.   
For knot A, we used for G130M wavelength setting 1222 with FP-POS 1 and 2, and setting 1291 with FP-POS 3 and 4, 
for a total exposure time of 1929 seconds. For knot B,  setting 1222 with FP-POS 2 and 3, and 1291 with FP-POS 3 and 4,
was used for a total exposure time of 1869 seconds. For G160M, both knots were observed with settings 1577 (FP-POS 2 and 3) and 
1611 (FP-POS 3 and 4) for a total exposure time of 2558 seconds per knot. All observations were obtained through the primary 
science aperture (PSA) using 'mirror A'. See Fig. \ref{fig:fig1} for the location of the COS apertures on \haro .

\begin{figure*}  
\centering
\includegraphics[angle=0,scale=0.5]{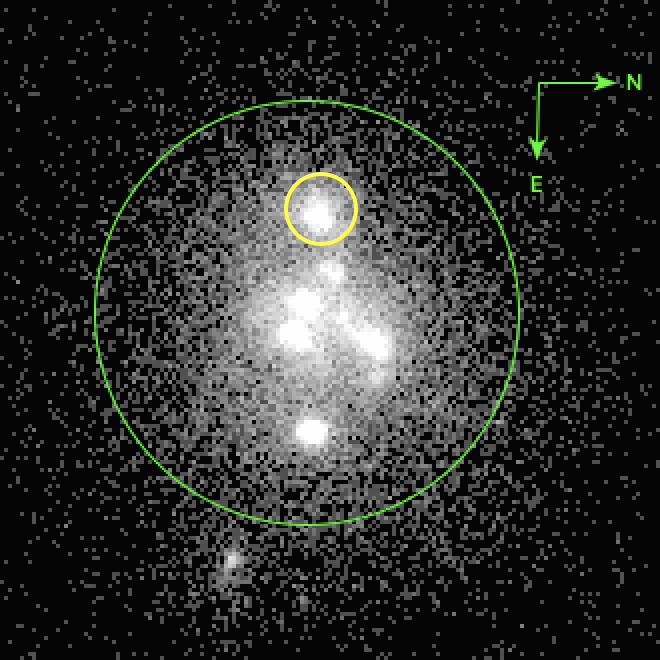}
\includegraphics[angle=0,scale=0.5]{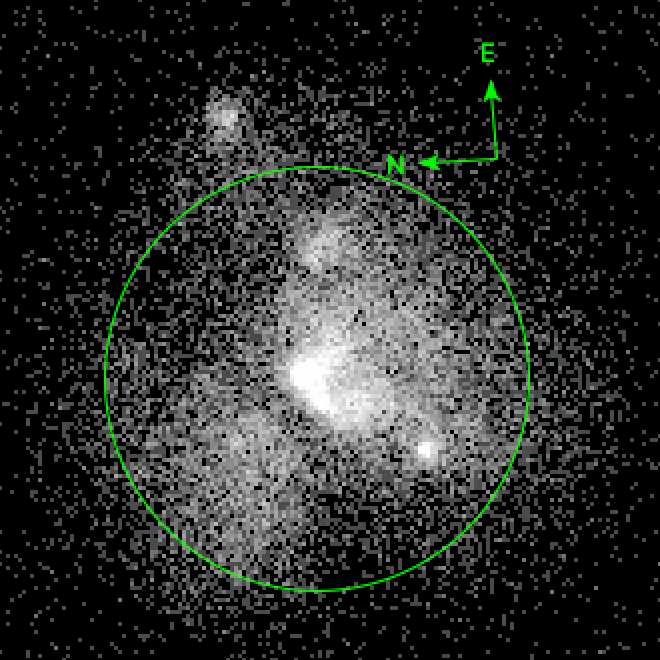}
\includegraphics[angle=0,scale=0.5]{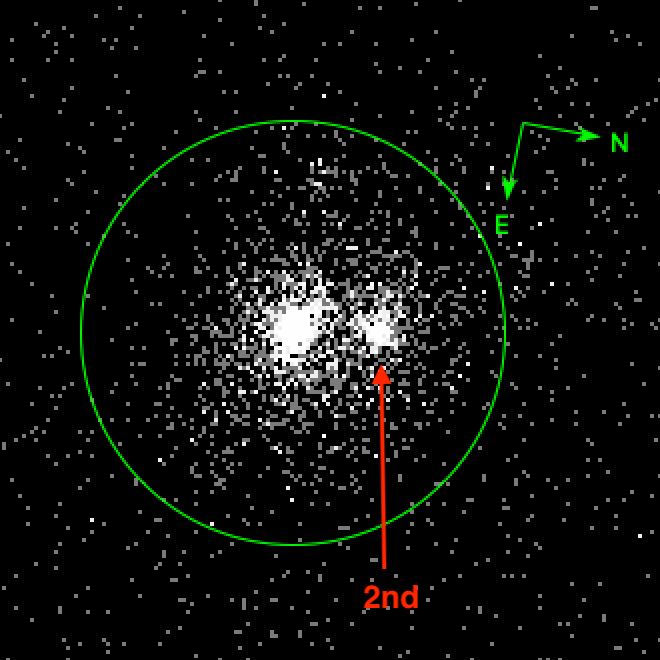}
\caption{
COS/NUV acquisition images for (from left to right) knot A, B and C, with logarithmic intensity scaling. The position of the nominal COS aperture ($\diameter = 2.5\arcsec$) is show as a green circle. The orientations on sky are indicated by the green compasses in the upper right corners. The dispersion direction for the FUV spectra is from left to right.  The images of knots A and B were obtained with the primary science aperture (PSA) and 'mirror A', while knot C (from GO 13017) was obtained with with PSA and 'mirror B', and shows 
the characteristic (for mirror B) secondary ghost image indicated by the red arrow. Note that this is just a characteristic of 'mirror B' and does not produce
a secondary FUV spectrum. 
Moreover, it should not be confused with the secondary source (green colored, $\sim 1\arcsec$ W of knot C)  in aperture C in Fig. 1, which is a bona fide star cluster. 
For knot A (left) the yellow circle indicates a source discussed in Sect. 4.} 
\label{fig:acq}
\end{figure*}

The spectral resolution of COS is degraded for extended sources. 
To mitigate this spectral degradation, we restricted the spacecraft ORIENT range for knots A and B such that each of the sources would have its minimal spatial
extent in the dispersion direction. For knot C, which consists of a single   point-like source, this is not
an issue, and the (archival) observations were obtained at  a {\tt PA$\_$APER} (the position angle on sky, measured from N to E, 
of the cross dispersion axis, i.e. y-coordinate in the detector frame) of {\tt 258.6} degrees. 
To get the position angle of the dispersion direction in the sky frame,
90 degrees is added to yield 348.6  degrees. 
In order to determine suitable ORIENT ranges for knots A and B we used available  ACS/SBC/F140LP far UV images from program 9470 (PI Kunth)
and ACS/HRC/F220W near UV images from program 10575 (PI \"Ostlin). For knot A, the COS observations were obtained at  {\tt PA$\_$APER\,=\,-90.0} 
so the dispersion direction is 0$\degr$, i.e. directed to N. For knot B, the observations were done at {\tt PA$\_$APER\,=\,94.2} so the dispersion direction is 184$\degr$,  approximately south, see Fig. \ref{fig:acq}.

Individual observations were reduced and extracted with \textsc{Calcos} version 3.3.4; extracted spectra were combined using custom splicing scripts.  Spectra were first rebinned by a factor of six because of the oversampling of the spectrograph, and to Nyquist sample the nominal line-spread function.

\subsection{MUSE data}
ESO/VLT/MUSE: \haro\ has been observed with the MUSE integral field spectrometer  \citep{bacon2010}  on ESO/VLT in the
extended  setting covering a wavelength $\sim 4650 - 9500$\AA . The reduction of these data is described in \citet{menacho2019}
and we here summarize the most important steps.

The sky subtraction was performed in two steps. In the first step, a sky mask was created by the MUSE pipeline "muse\_scipost" from the 20 percent darkest area of each science exposure.
These sky regions enclose, however, also  areas in the halo with real but faint \ha\ and [OIII]5007 emission, and as 
consequence produced an oversubtraction of the sky. Thus, in a second 
step we customized the sky mask by removing these emission areas.
The  sky residuals are $<2 \cdot10^{-20}$ \egs\ per pixel, or maximum $10^{-18}$ \egs \ over a COS aperture, while \ha\ and \hb\ line  fluxes are 4 to 5 orders of magnitude 
greater. Hence, uncertainties in the sky level do not incur significant uncertainties in the \ha / \hb \ values or the derived extinctions.
The \ha\ and \hb \ spectra were corrected for underlying stellar absorption with the pPXF package \citep{cappellari2017} using the 
 the MILES stellar spectra library \citep{miles1,vazdekis2010}.
We use this data to measure optical emission line fluxes in the three knots.

\begin{deluxetable}{lrrrl}[b!]
\tablecaption{Measuremernts for knots A, B and C. \label{tab:results}}
\tablecolumns{5}
\tablenum{1}
\tablewidth{0pt}
\tablehead{
\colhead{quantity} &
\colhead{knot A} &
\colhead{knot B } &
\colhead{knot C} &
\colhead{unit/comment} 
}
\startdata
$F_{\rm H\alpha}$ & 2.27 & 5.19 & 1.40& ($10^{-13}$erg/s/cm$^2$) \\
$v_{H\alpha}$ &+57 & -24& -54& (km/s) \\
$v_{star}$ &+51 & -24& -21& (km/s) \\
$\sigma_{H\alpha}$ & 71.1 & 89.2 & 64.2 & (km/s) \\
$EW(H\alpha)$ & 581 & 827 & 153  & (\AA)  \\
$\rm [OIII]/H\beta$ & 4.47 & 4.06 & 3.19 & 5007/$H\beta$ \\
$\rm [SIII]/[SII]$ & 3.18 & 3.63 & 1.55 & {\footnotesize (9531+9069)/(6717+6730)}\\
(\ha/\hb)$_{raw}$ & 3.52 & 4.38 & 4.24 & observed \\
(\ha/\hb)$_{corr}$ & 3.48 & 4.33 & 4.19 & corrected for MW ext\\
$E(B-V)_{n}$ & 0.18 & 0.38 & 0.35 & from (\ha/\hb)$_{corr}$  \\
$\beta$ & -1.73 & -0.86 & -1.89  & UV continuum slope \\
$F_{\rm Ly\alpha}$ & 0.349 & 0.656 & 1.515 & ($10^{-13}$erg/s/cm$^2$)\\
$W_{Ly\alpha}^{red}$ &302 & 338& 195 & (km/s)  {\small FWHM} red peak \\
$\Delta v_{Ly\alpha} $ & 532 & 409 & {\it (400}\,$^\dagger$) & (km/s) peak separation \\
$A_{red}$ & 1.5 & 2 & 2.7 & \lya\ red peak asymm\\
$f_{esc}^{Ly\alpha}$ & 0.012 & 0.0065 & 0.059 & \lya\ escape fraction \\
${\mathcal R}$ & 0.072 & 0.29 & 2.0 & 'ext corrected' $f_{esc}^{Ly\alpha}$  \\
\enddata
\tablecomments{ $F_{\rm H\alpha}$ is the \ha\ flux, corrected for foreground reddening. 
$v_{H\alpha}$ and $v_{star}$ are velocities derived from  \ha\ emission, and UV photospheric absorption lines, respectively, and 
are with respect to the systemic velocity of 6194 km/s. Uncertainties are $<3$ km/s, and $< 10$ km/s, respectively. The \ha\ line width,
$\sigma_{H\alpha}$, 
has uncertainties $<4$ km/s, and $EW(H\alpha)$ is accurate to $<3$\%. 
(\ha/\hb)$_{raw}$ is the observed line ratio, and (\ha/\hb)$_{corr}$ the one corrected for foreground extinction
(accurate to $\sim 0.03\%$). 
$F_{\rm Ly\alpha}$ is the \lya\ flux, corrected for foreground extinction; the uncertainties are dominated by the 
intrinsic COS calibration and the continuum determination and is accurate to $\sim10\%$. 
 $W_{Ly\alpha}^{red}$ has uncertainties of $\sim10\%$ while peak separations have uncertainties of 
 $\sim 20$km/s. Red peak asymmetries are accurate to $\sim5\%$. 
 The uncertainties on $f_{esc}^{Ly\alpha}$ and ${\mathcal R}$ are dominated by the \lya\ 
 accuracy and is hence $\sim10\%$.
$\dagger$: we report the peak separation in knot C,  assuming the small blue bump is the blue peak, although it is more
likely due to a P-Cygni profile.  See main text for further details.}
\label{table1}
\end{deluxetable}

\begin{deluxetable*}{lrrrll}[b!]
\tablecaption{Results for absorption lines of knots A, B and C. \label{tab:aod}}
\tablecolumns{6}
\tablenum{2}
\tablewidth{0pt}
\tablehead{
\colhead{quantity} &
\colhead{knot A} &
\colhead{knot B } &
\colhead{knot C} &
\colhead{unit} &
\colhead{comment} 
}
\startdata
$v_{min}$ & --159\,$^{+17}_{-1}$ & --85\,$^{+17}_{-17}$ & --139\,$^{+17}_{-51}$ & km/s & \sit\ \\
$v_{int}$ & --228\,$^{+1}_{-1}$  & --137\,$^{+17}_{-1}$ & --169\,$^{+1}_{-1}$ & km/s  & \sit\ \\
$v_{5\%}$ & --57\,$^{+1}_{-17}$  & 51\,$^{+17}_{-17}$ & 37\,$^{+1}_{-17}$ & km/s  & \sit\ \\
$v_{95\%}$ & --468\,$^{+17}_{-17}$  & --377\,$^{+17}_{-51}$ & --425\,$^{+34}_{-34}$& km/s & \sit\ \\
$W_{90\%}$ & 411\,$^{+17}_{-34}$  & 428\,$^{+51}_{-34}$ & 462\,$^{+34}_{-34}$ & km/s & \sit\ \\
$f_{c, v_{min}}$ & 0.95\,$^{+0.05}_{-0.04}$	 & 0.96\,$^{+0.04}_{-0.14}$ & 0.55\,$^{+0.05}_{-0.04}$&   &\sit\  \\
$f_{c, v_{sys}}$ & 0.32\,$^{+0.06}_{-0.10}$	 & 0.87\,$^{+0.13}_{-0.18}$ & 0.51\,$^{+0.02}_{-0.07}$&   &\sit\  \\
 $\log (N_{\rm SiII}, v_{min})$ & 13.0\,$^{+\infty}_{-0.2}$& 12.8\,$^{+\infty}_{-0.2}$ & 12.4\,$^{+0.1}_{-0.1}$  &  cm$^{-2}$/km s$^{-1}$  & \sit \\
 $\log (N_{\rm SiII}, v_{sys})$ & 12.6\,$^{+\infty}_{-0.4}$& 12.6\,$^{+\infty}_{-0.3}$ & 12.1\,$^{+0.1}_{-0.1}$  &  cm$^{-2}$/km s$^{-1}$  & \sit \\
 $\log(\Sigma f_c N_{\rm SiII})$ & 15.2& 15.3 & 14.7  &  cm$^{-2}$ & \sit \\
$f_{c,  v_{min}}$ & 0.93\,$^{+0.07}_{-0.04}$ & 1.0\,$^{+0}_{-0.22}$ & 0.94\,$^{+0.06}_{-0.1}$&  & \sif \ at $v_{min}$ for \sit \\
$f_{c,  v_{sys}}$ & 0.40\,$^{+0.19}_{-0.07}$ & 0.77\,$^{+0.23}_{-0.11}$ & 0.98\,$^{+0.02}_{-0.15}$&  & \sif \ at $v_{sys}$ \\
 $\log(N_{\rm SiIV}, v_{min})$ & 12.8\,$^{+\infty}_{-0.4}$  & 12.3\,$^{+\infty}_{-0.1}$  & 12.2\,$^{+0.2}_{-0.1}$  &  cm$^{-2}$/km s$^{-1}$  & \sif \ {at $v_{min}$ for \sit }\\
 $\log(N_{\rm SiIV}, v_{sys})$ & 13.6\,$^{+\infty}_{-1.7}$  & 12.6\,$^{+\infty}_{-0.5}$  & 12.0\,$^{+0.2}_{-0.1}$  &  cm$^{-2}$/km s$^{-1}$ & \sif \ {at $v_{sys}$  }\\
$\log(\Sigma f_c N_{\rm SiIV})$  & 15.0 & 14.8 & 14.8 & cm$^{-2}$ & \sif \\
\enddata
\tablecomments{The first 4 lines give kinematic properties for the \sit\  lines, where $v_{min}$ is the velocity at the deepest absorption,
$v_{int}$ the absorption weighted average velocity, with respect to the \ha\ velocity for each knot.  Summing all the \sit\ absorption,  $v_{5\%}$ and $v_{95\%}$  is the velocity where 
5\% and 95\% of the total absorption is encompassed, respectively.  $w_{90\%} = v_{95\%} - v_{5\%}$, is our measure of the total \sit\
absorption velocity width. The sixth and 11th rows give the \sit\ and \sif\ covering fractions, respectively, both at $v_{min}$ for \sit , while row
7 and 12 give $f_c$ for the systemic velocity of each knot.
The errors represent the confidence intervals based on Monte Carlo simulations \citep{rivera-thorsen2015,rivera-thorsen2017}.
For the column densities, the unit is cm$^{-2}$ per km s$^{-1}$, and the upper limit is sometimes unconstrained (represented by $+\infty$). $\Sigma f_c N_{\rm [SiII]}$ is the sum of the covering fraction times the column density and velocity bin width for \sit\ and  $\Sigma f_c N_{\rm [SiIV]}$ the same for \sif . Given the very asymmetric confidence intervals for the column densities, it is not meaningful to present uncertainties on these quantities, which should be regarded as quite approximative. }
\label{table2}
\end{deluxetable*}

\bigskip

\section{Results}
Here we describe the results and measurements done for each of knots A, B and C.

\subsection{MUSE}

\begin{figure*}  
\centering
\includegraphics[angle=0,width=18cm,trim=0cm 0 0cm 0]{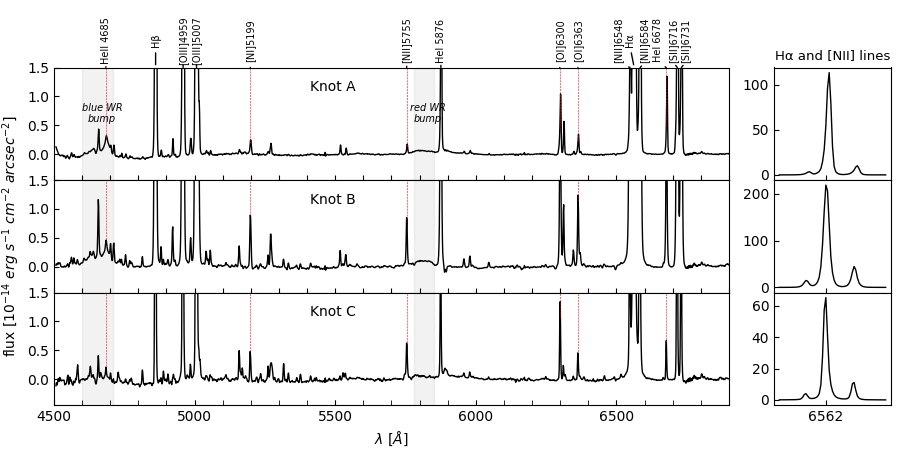}
\caption{MUSE spectra for the 3 knots, matched to the COS apertures. The abscissa shows the rest wavelenth over the range 4500-6900\AA .
Prominent lines and the location of the WR bumps are indicated. On the right, we show zoom-ins for the regions around \ha\ and \nii .} 
\label{fig:muse}
\end{figure*}

We used the COS acquisition images to find the exact location of the spectroscopic apertures in the MUSE image plane. Since the throughput of the  $\diameter = 2.5\arcsec$ 
COS aperture is not a top-hat function, but subject to vignetting and also transmits some flux from $ 1.25\arcsec < r < 2\arcsec$, we multiplied the
MUSE spectral cube with the COS aperture vignetting function \citep{goudfrooij2010}. Then 1D MUSE spectra were extracted by integrating the flux 
over the vignetted aperture. Hence
these MUSE extractions  spatially matches the COS spectra for each of the knots, and were used
to measure emission line fluxes of \ha , \hb , \oiii , \sii\ and \siii . To determine the  \ha\ velocity and line width,  we
fitted the \ha\ profile with a single Gaussian, and for the line width ($\sigma_{H\alpha} = FWHM/2.355$) subtracted 
the MUSE line spread function in quadrature \citep{menacho2019}.
The line fluxes were corrected for foreground Milky Way reddening of $E(B-V)=0.010$ \citep[NED\footnote{NASA/IPAC Extragalactic Database: http://ned.ipac.caltech.edu},][]{schlafly2011} . The statistical uncertainties on the line fluxes are $\lesssim 0.03\%$ for \ha , \hb, \oiii ; and $\lesssim 0.5\%$ for the \sii\ and \siii\ lines. 
Various line ratios (all accurate to $<1\%$) are given in Table \ref{tab:results}.
For $[SIII]/[SII] = ([SIII]9531+9069) / ([SII]6717+6730)$ the 9531 transition was not in the wavelength region 
observed and we used  $[SIII]9532 = 2.469\cdot[SIII]9069$  from \citet{agn2}.  
The total uncertainties are dominated by the 
absolute MUSE flux calibration (good to a few \%). The MUSE spectra for the three knots are shown in Fig. \ref{fig:muse}

From the corrected  \ha/\hb\ ratio we derived the nebular reddening expressed as $E(B-V)_{n}$ by adopting the \citet{fitzpatrick1999} extinction law
(with $R_V=3.1$ and $k_{H\beta} - k_{H\alpha}=1.23$) and an intrinsic  \ha/\hb\ ratio of 2.84 \citep[approriate for case B $T_e\gtrsim 10$K,][]{agn2}.

The results are presented in Table \ref{tab:results}. All three knots have appreciable nebular extinction, but both knots A and C have 
 blue UV continuum slopes suggesting low stellar reddening (see section 4.2 for further discussion).    

Knot B has the highest \ha\ flux, the highest $EW(H\alpha)$ and highest reddening. This is obviously where the ionizing photon production rate is highest. Knot A also appears very young and contains many massive young star clusters 
\citep{adamo2010}. Both knots A and B have high ionization as  judged by \oiii/\hb\ and \siii/\sii . Knot C has lower EW(\ha) and ionization
but a higher electron temperature and lower metallicity \citep[]{james2013,menacho2021}. While compact, with $r_{\rm eff} \approx 40$\,pc \citep{ostlin2015}, it is 
clearly more extended than  the compact star clusters making up knots A and B, and more resembling of a nuclear star cluster
and likely has experienced an extended star formation history.

\subsection{ \lya\ emission}
\begin{figure*}  
\centering
\includegraphics[angle=0,scale=0.745,trim=0.75cm 0 0 0]{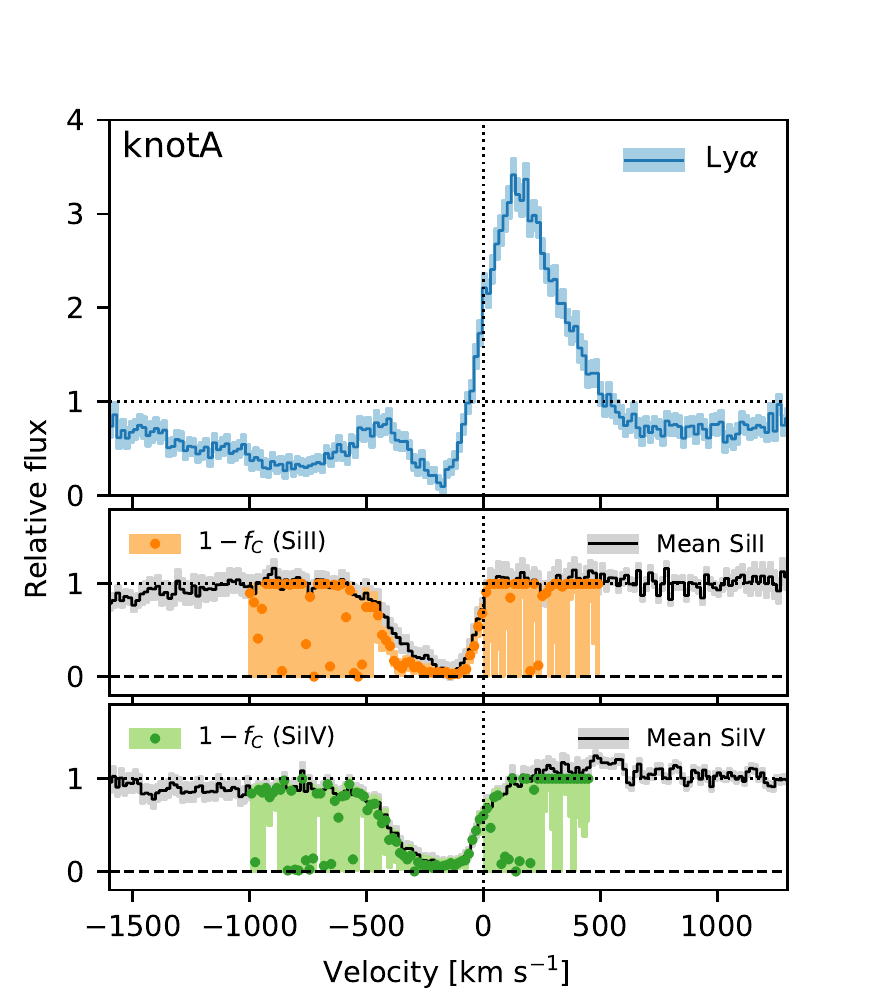}
\includegraphics[angle=0,scale=0.745,trim=0.7cm 0 0 0]{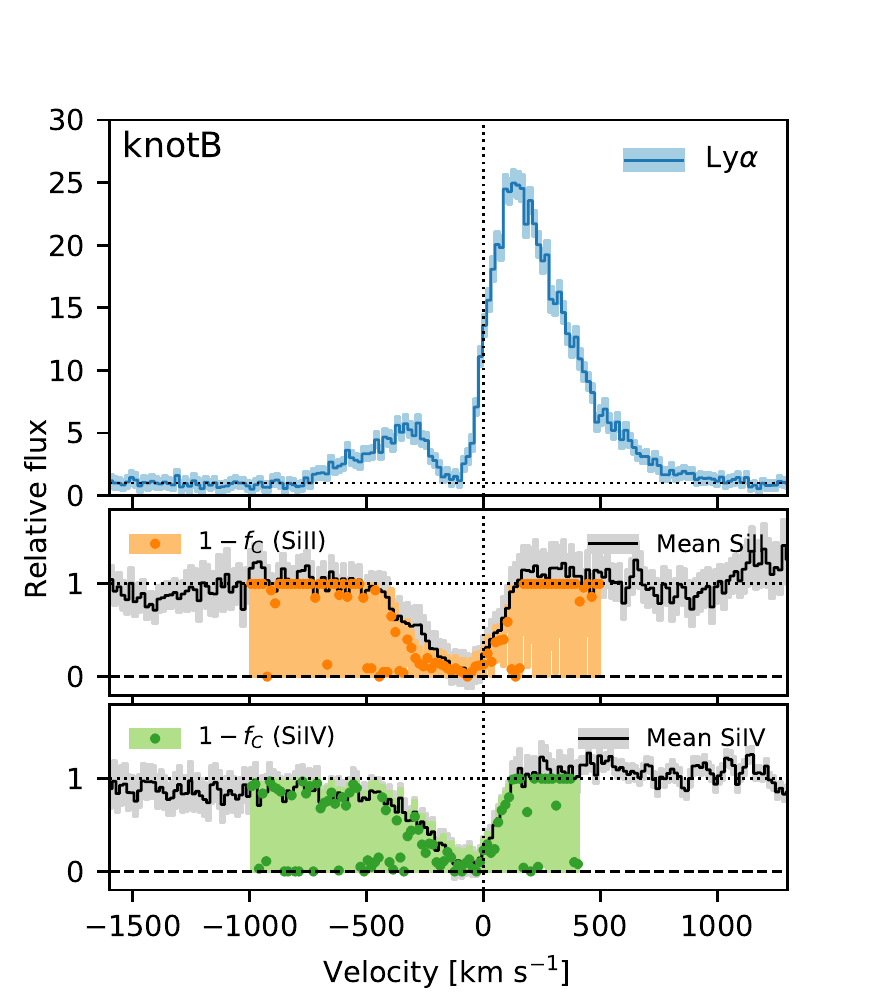}
\includegraphics[angle=0,scale=0.745,trim=0.7cm 0 1.2cm 0]{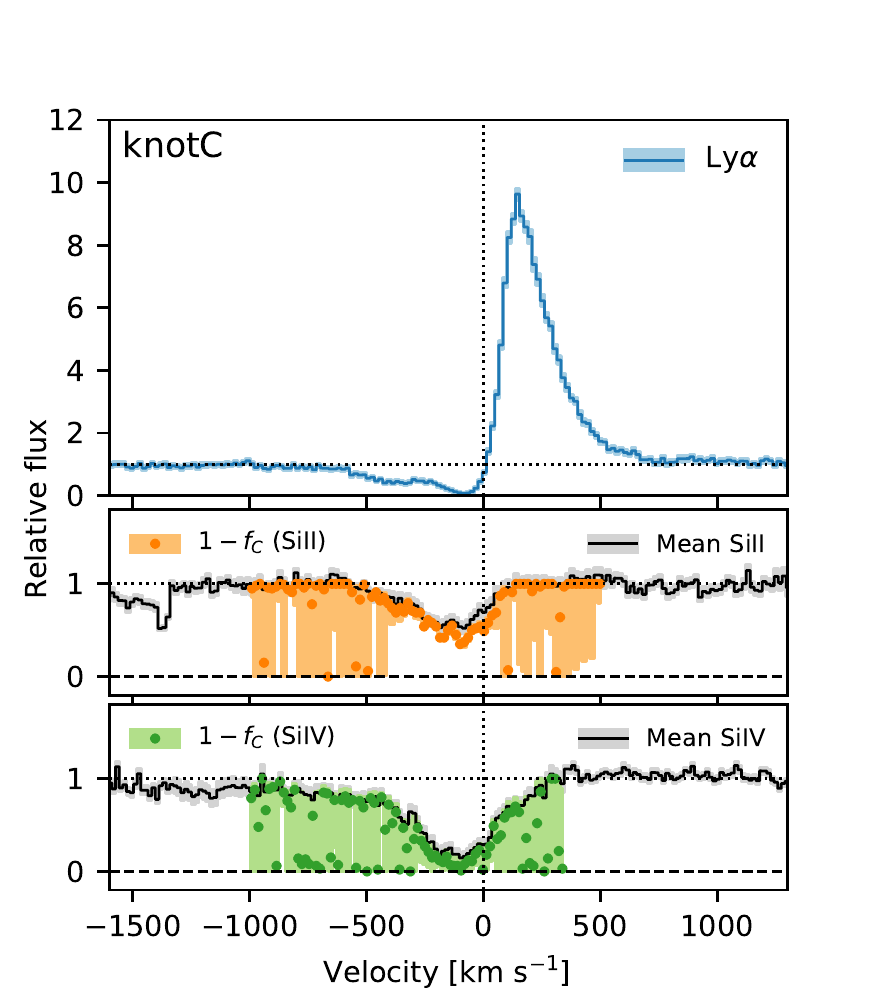}
\caption{\lya\ and ISM line spectra of (from left to right) knots A, B and C. The upper panels shows the \lya\ profiles
(normalized such that the continuum at 1230\AA\ is 1). 
The middle panels shows the average profile 
of the  \sit\ absorption lines in black, normalized to one in the continuum, and in orange '$1-f_c$' (one minus the covering fraction, i.e. the {\em non} covering fraction) resulting from the AOD analysis.
 The bottom panel shows the average \sif\ profile in black, and in green   '$1-f_c$' for \sif . The shaded regions in all panels represent the $1\sigma$ uncertainties, and for '$1-f_c$' the confidence 
 intervals as described in \citet{rivera-thorsen2015}.  The velocity zero point  is that derived from \ha \ for each knot.} 
\label{fig:aod}
\end{figure*}

In addition to the previously known emission from knot C \citep{rivera-thorsen2017} we detect prominent, though fainter, \lya\
also from knots A and B. The \lya\ profiles are presented in Fig. \ref{fig:aod} (where the \lya\ flux is given in units of the local continuum, 
normalized to 1 at rest wavelength $\lambda_0=1230$\AA ). 
The three knots differ in their \lya \ properties. The \lya\ profile for knot C  \citep[previously published in][]{rivera-thorsen2017}  shows
a strong red peak and a hint of a blue peak, which however is more resemblant of a P-Cygni profile. 
Knots A and B  show distinct blue secondary peaks, and in addition broad continuum
absorption near \lya . We measure the velocities of the red and blue peaks,
the width of the red peak,
and determine the peak separation (see Table \ref{table1}, where we also include the asymmetry of the red peak as defined by
 \citet{kakiichi2019} for future reference). For the purpose of measuring the \lya\ escape fraction locally from the three knots, we integrated
the \lya\ flux using the local continuum located at approximately  $\pm 6$\AA , as we want to evaluate the amount of \lya\ photons that are directly emerging
from each knot.  Hence not all broad continuum absorption is  accounted for.  The flux uncertainties are dominated by the calibration accuracy of 
COS which which is quoted to be 5\% \citep[][ although the accuracy for extended sources is somewhat worse]{cos-ihb} and the continuum placement,
and is overall estimated to be $\lesssim 10$\%. The \lya\  fluxes were corrected for  MW reddening with the same recipe as above \citep{schlafly2011,fitzpatrick1999} resulting in an upwards correction of 9.8\%. 
The effective resolution of COS depends on the 
spatial structure of the source, and since \lya\ is in general extended the resolution is degraded to $R\sim2000$ 
\citep[FWHM 150 km/s][]{rivera-thorsen2017}, but may differ somewhat from source to source depending on the structure of the emission.
In any case, the non-zero \lya\ emission seen at zero velocity in knot A and B is likely a product of the limited spectral resolution, and
not a signature of \lya\ photons at line center directly escaping without being scattered \citep[see Fig. 3 in][]{verhamme2015}.

Knot A has a broad red \lya\ peak, and a peak separation of $>500$ km/s, suggesting  scattering at relatively high \hi\ column densities. 
Knot B has a similarly broad red peak, but a smaller  peak separation compared to knot A. Hence
the  \hi\ optical depth is likely comparable to or somewhat smaller than in knot A.
Knot C has the narrowest \lya\ profile of the three knots, 
suggesing a lower \hi\ scattering column density \citep{verhamme2006,verhamme2015}.

\subsection{ \lya\ escape fractions}

 We use the MUSE \ha\ flux and reddening together with the COS \lya\
flux measurements to calculate the \lya\ escape fraction for each knot within the COS aperture according to: 

$$ f_{esc}^{Ly\alpha} = F_{Ly\alpha} / (8.4 \cdot F_{H\alpha} \cdot 10^{0.4\cdot E(B-V)_n \cdot k_{6563}} )$$ 																							
\citep[see][]{hayes2011} where 8.4 is the Case B \lya /\ha\ value for $T\gtrsim10^4$K and $n_e\sim 100$ cm$^{-3}$\citep{brocklehurst1971,hayes2019}, $E(B-V)_n$ is the nebular reddening derived from \ha/\hb\ and $ k_{6563}=2.32$ is the relative extinction coefficient at \ha . The derived
escape fractions for knots A, B and C are presented in Table 1. Knot A and B have low escape fractions ($\lesssim1\%$) while knot C has
$ f_{esc}^{Ly\alpha}=6$\%. 
Since, due to resonant scattering, a fraction of the \lya\ photons originating in the 
knots may escape  outside the aperture rather than being absorbed by dust, the actual escape fractions are likely  higher than the
numbers given above. 

We also
compute ${\mathcal R}$ as in \citet{ostlin2009}: 

$$ {\mathcal R} = \frac{1}{8.4}\frac{F_{Ly\alpha}}{F_{H\alpha}} \cdot 10^{0.4\cdot E(B-V)_n \cdot (k_{6563} - k_{1216})} $$ 																

which measures how the extinction corrected \lya /\ha \ ratio deviates 
from the  prediction from a simple dust screen geometry where \lya\ and \ha\ are
exposed to the same amount of dust.  $ {\mathcal R}$ can be viewed as the extinction corrected escape fraction:  If \lya\ is only attenuated by
dust and subject to the same dust optical depth as \ha , then  	${\mathcal R}$ should be equal to 1. 
We take  $(k_{6563} - k_{1216} =8.6) $ from the starburst law of  \citet{calzetti2000} and the ${\mathcal R}$ values  are given in Table \ref{tab:results}. 
In contrast with \ha , the extinction correction for \lya\ is sensitively dependent on the chosen extinction 
law: the  \citet{fitzpatrick1999}, \citet{cardelli1989} and \citet{calzetti2000} laws all give similar results (to within 20 percent), 
while  the SMC law \citep{prevot1984} give ${\mathcal R}$ values  that are several times higher (0.2, 2, and 12 for knots A, B and C, 
respectively). This suggests that the SMC law (derived for stars) does not provide a fair estimate of the attenuation suffered by \lya\ photons.

The uncertainties on  $f_{esc}^{Ly\alpha}$ and $ {\mathcal R}$ are dominated by the \lya\ flux uncertainties, 
and the accuracy of the COS vignetting function (i.e. how well matched the effective \lya\ and \ha\ apertures are).

\subsection{Broad HI absorption and Voigt fitting}

\begin{figure}  
\centering
\includegraphics[angle=0,width=9.4cm,trim=1.4cm 0 0cm 0]{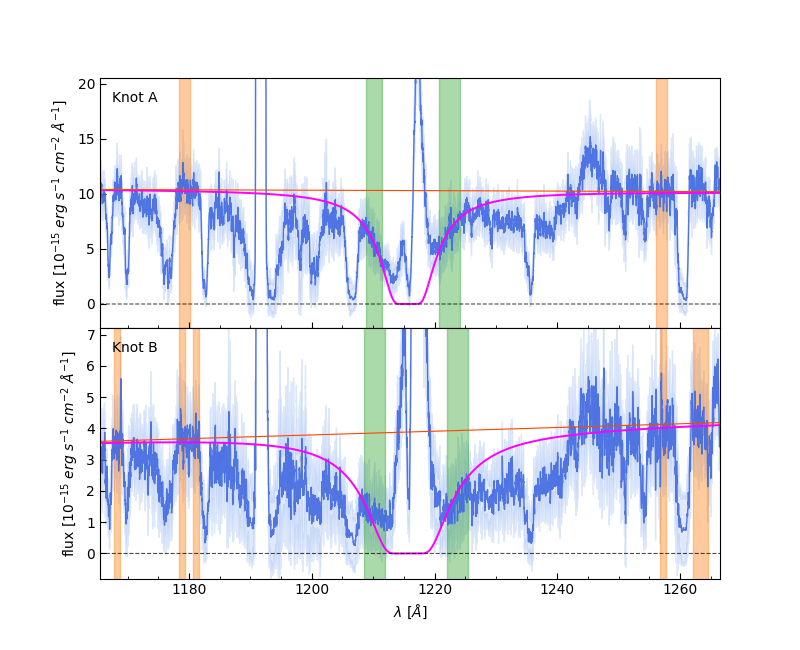}
\caption{Voigt profile fitting of the \lya\ absorption wings in knot A (upper) and B (lower). Orange shading represent the regions
for continuum normalization, and green shading the region for fitting of the Voigt profiles. These fits provide column densities of 
20.7 and 21.0 cm$^{-2}$ for knots A and B, respectively. } 
\label{fig:voigt}
\end{figure}

Both knots A and B show a decreased continuum level close to \lya  . 
In order to assess whether this could be due to damped neutral 
hydrogen absorption, we performed Voigt profile 
fitting with a single component to the absorption wings of \lya . 
 We fix the absorption velocity for
the atomic gas to that measured from the \sit\ 1260 absorption line, but
leave the doppler parameter and column density free.  Because of many
(sometimes blended) absorption components in the vicinity of \lya\ (N{\sc v}
PCyg, \sit\ ISM lines, and the same ISM features in the MW), we pay
careful attention to the continuum placement, isolating regions we
believe to be clean.  On the red side we adopt a region between the
emission peak of the N{\sc v} 1240 Pcyg and the \sit\ 1260 ISM line, while on
the blue side we adopt a narrow region between the stellar C{\sc iii} 1175
line in Haro 11 and the Si{\sc iii} 1206 \AA\ absorption line in the Milky Way
(shifted back to 1182 \AA\ in the restframe representation of
Figure~\ref{fig:voigt}).   For the Voigt profile fitting itself, we also adopt
regions that we believe to be free from discrete interstellar absorption
lines, and the extended blue wing of the N{\sc v} PCyg, which when blending
with the \lya\ absorption may appear relatively flat as an artificially
reduced continuum  at $\sim$1230 \AA .

The results are shown in Fig. \ref{fig:voigt}. 
For knot A we find a column density of  $\log(N_{\rm H\,I})=20.7$, and for 
 knot B,  $\log(N_{\rm H\,I})=21.0$.  
 No sign of damping wings is  seen in knot C, and we could not produce a viable fit.

\begin{figure}  
\centering
\includegraphics[angle=0,scale=0.7,trim=0.2cm 0 0 0]{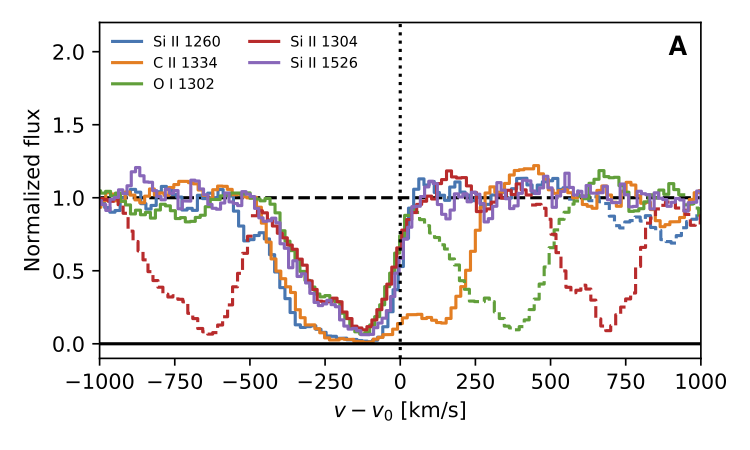}
\includegraphics[angle=0,scale=0.7,trim=0.2cm 0 0 0]{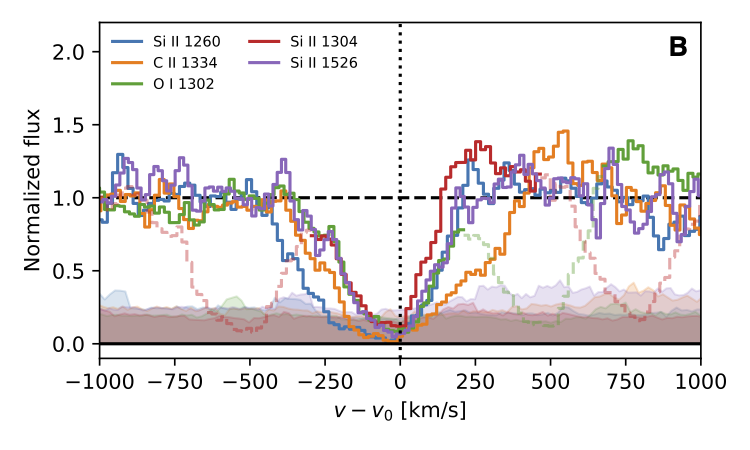}
\includegraphics[angle=0,scale=0.7,trim=0.2cm 0 0 0]{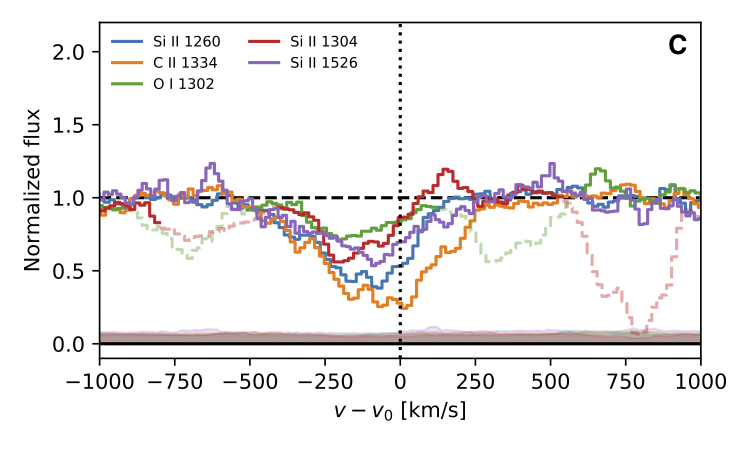}
\caption{Selected LIS lines on a  common velocity scale, for (from top to bottom) knot A, B and C (respectively). The velocity zero point for each knot is that determined from the
\ha\ 
emission line and given in  Table \ref{table1}. The LIS lines are color coded and shown as solid lines, while the dashed parts of the lines show adjacent features. 
E.g. the O\,{\sc i} 1302 transition (solid green) shows the neighbouring Si\,{\sc ii} 1304  line (dashed green, to the right), and vice versa (O\,{\sc i}\,1302 line shown as dashed red 
to the left of Si\,{\sc ii} 1304). The feature (dashed red) seen on the right side of Si\,{\sc ii} 1304 (solid red) in all panels is due to C\,{\sc ii} 1334 from the Milky Way. 
The shaded regions close to zero intensity shows the error spectrum, but is to small to be seen in knot A.} 
\label{fig:lis}
\end{figure}

\subsection{Stellar velocities} \label{vstar}
We used photospheric lines from \ciii\ (1247 and 1776 \AA) to measure the velocities of the UV-bright stellar 
components, i.e. the ionizing sources. 
For knot A and B, these agree with the velocities inferred from \ha , while  for knot C, the stellar velocities  
are 33 km/s higher, see Table \ref{table1}. Hence, the \ha\ emitting gas towards knot C is on average 
blueshifted with respect to the ionizing stars, but the difference is just one fifth of the \ha\ line
width (FWHM$_{H\alpha}\,=2.355\, \sigma_{H\alpha}=150$ km/s, see Table \ref{table1}) and does not affect any of the analysis
in this paper.

\subsection{ISM absorption lines, covering fraction, column densities} \label{AOD}

The spectra show the presence of metal absorption lines arising both from the Milky Way and \haro ,
notably several transitions of singly and triply ionized silicon. Since the ionization potential for silicon (8.2 eV) 
is lower than for hydrogen (13.6 eV), \sit\ primarily probes the part of the ISM where
hydrogen is neutral. Other probes of the neutral ISM in the COS spectra of \haro\ are \oi\ 1302 and 
\cii\ 1334 \AA. We show the absorption profiles of these so called Low Ionization State (LIS) lines in Fig. \ref{fig:lis}.
Conversely, to triply ionize silicon, photons with energy 33.5 eV are required and  \sif\ hence traces
highly ionised gas (similar to \oiii ).

We analyse the UV absorption line spectrum using the same methods as in \citet{rivera-thorsen2015,rivera-thorsen2017} 
based on the apparent optical depth (AOD) method \citep[as implemented in][]{jones2013} utilising three Si\,{\sc ii} lines (1260, 1304, 1526 \AA)
with the same lower state but have different oscillator strength, to solve for the 
column density and covering fraction as a function of velocity. The use of three lines with different strength also improves the 
estimates of column densities if the lines are saturated. For the warm  ionised gas we perform the same analysis using two
Si{\sc iv} transitions (1393, 1402 \AA). The results from knot C have been presented in \citet{rivera-thorsen2017}.  
Here we perform the same analysis  for knots
A and B (and for consistency redo the one for knot C, using the updated \ha\ velocity in Table \ref{table1}). 

In each velocity bin ($i$), the residual intensity (in units of the continuum intensity) is given by:
$$ \frac{I_i}{I_0}= 1- f_{c,i} (1-e^{-\tau_i}) $$

where $f_{c,i}$ is the covering fraction, and $\tau_i$ the optical depth, given by:

$$ \tau_i = f \lambda \frac{\pi e^2}{m_e c} N_i = f \lambda \frac{N_i}{3.768\times10^{14}} $$

where $f$ is the oscillator strength for the transition in question and $N_i$ the column density 
for the given velocity bin and ion. The fact that the \sit\ line with the highest oscillator strength 
(1260 \AA) has in general a deeper profile than the other two, indicates that the latter have lower
optical depth and hence are not completely saturated, see Fig. \ref{fig:lis}.

	Since absorption occurs from the bright compact UV emission, the spectral resolution is 
only moderately degraded, \citet{rivera-thorsen2017} estimate  $R\approx 10\,000$ or 30 km/s for knot C, but we note that knots A and B have 
more complicated UV morphologies, and hence possibly somewhat lower resolution. 
We investigated the COS/NUV acquisition images (see Fig. \ref{fig:acq}) and collapsed them in the along the cross dispersion direction to find the FWHM 
(and half light width, HLW) in the dispersion direction and obtained 0.34\arcsec\ and 0.31\arcsec\ (HLW $=$0.34\arcsec and 0.53\arcsec) for knots A and B, respectively. 
 
\begin{figure*}  
\centering
\includegraphics[angle=0,scale=0.7,trim=0.2cm 0 0 0]{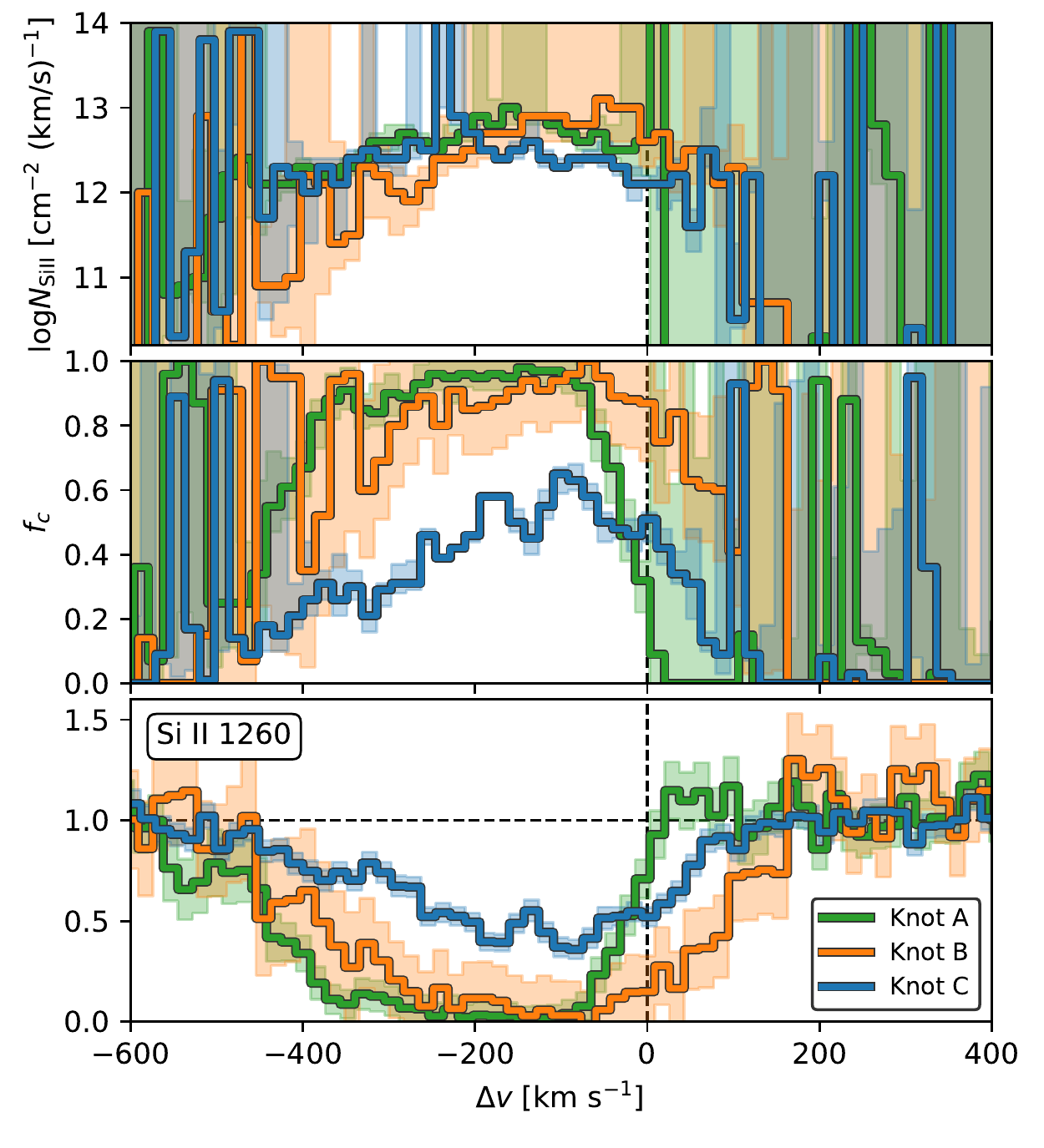}
\includegraphics[angle=0,scale=0.7,trim=0.2cm 0 0 0]{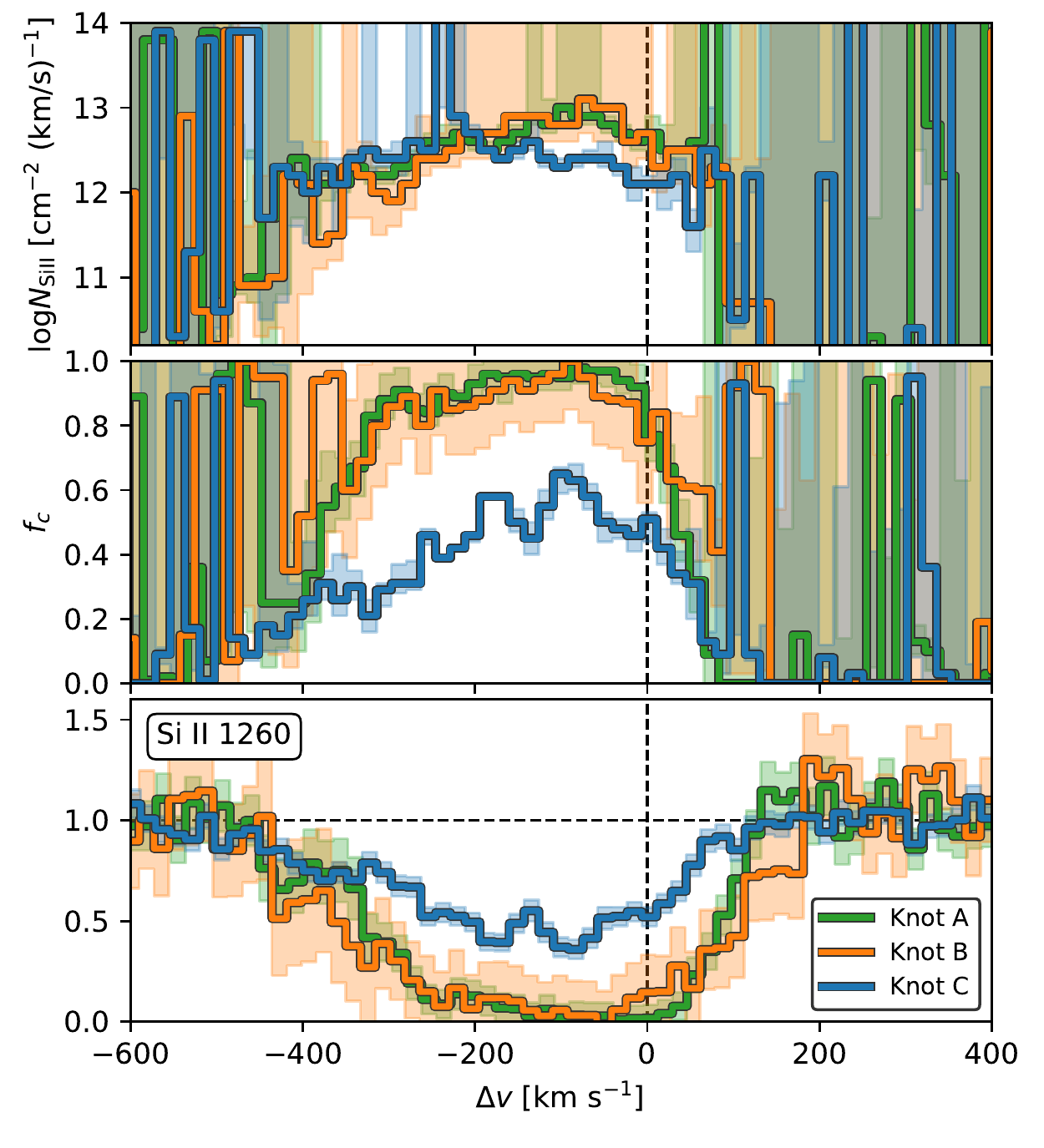}
\caption{The upper panels shows the comparison of the \sit\ column densities derived from the AOD analysis of 
knots A (green), B (orange), and C (blue).  The middle panels shows the inferred \sit\ covering fractions. The bottom
panels show the profiles of the strongest \sit\ transition ($\lambda_{\rm rest}=1260$\AA ). {\bf Left:} The reference velocity is here the
individual values for each of knots A, B, and C. {\bf Right:}  The reference 
velocity for all knots is here the systemic velocity of  knot C.} 
\label{fig:grid}
\end{figure*}

The COS/FUV beam is considerably aberrated and 
point sources have FWHM  in the cross dispersion direction of 0.5-1.5\arcsec\ \citep[COS ISR 2018-08, ][]{isr2018-08}, and sources with FWHM\,$\ge$0.6\arcsec\ should be 
regarded as extended \citep[COS IHB,][]{cos-ihb}. Hence, the continuum spectral resolution of knots A and B should be only marginally degraded by the low surface
brightness wings. On the other hand, the spectral resolution at LP4 is degraded to approximately 13\,500 \citep[COS ISR 2018-07,][]{isr2018-07}. 
Hence a spectral resolution of $R\sim$10\,000 is a fair approximation for each of the knots. Inspection of some Milky Way lines (e.g. \sit\ 1260 and \cii\  1335)
indeed verifies that the spectra towards all three knots appear to have the same resolution. 
Our  velocity sampling of 17 km/s hence oversamples the effective resolution by a factor of $\sim2$ (i.e. approximately Nyquist sampling),
and is close to the thermal line width.
 The results of the AOD analysis  are shown in Figs. \ref{fig:aod},\ref{fig:grid} and Table \ref{tab:aod}. The derived covering fractions vary with velocity and 
 show a maximum close to the deepest \sit\ absorption.

The neutral ISM traced by Si\,{\sc ii}, which is sensitive to \hi\ gas with column density $\gtrsim 10^{17}$ cm$^{-2}$   \citep{rivera-thorsen2017},
displays a velocity width ($W_{90\%}$ encompassing 90\% of the absorption) of more than 400 km/s for all three knots. This gas is 
very optically thick to \lya\ but has $\tau_{LyC}\sim 1$ for a single velocity bin.
The deepest absorption is consistently seen at 
velocities that are blueshifted  relative to \ha . 
  Notably  the Si{\sc ii} profiles mimic the shape of the \lya\ profiles between the two peaks, a behavior
also seen in the LARS sample \citep{rivera-thorsen2015}, i.e. \lya\ intensity quickly drops at velocities where there is a
large covering of neutral gas optically thick to \lya .
The fact that we see neutral ISM absorption over a wide range of velocities and with varying (non unity) covering fractions demonstrates 
that we have an outflowing clumpy neutral medium along each sightline, rather than a single neutral shell \citep{rivera-thorsen2015}.
Knot A shows a flat  covering fraction distribution  for $v\le-70$km/s with $f_c >0.9$ in 12 consecutive velocity bins (i.e. over a range 200 km/s),
 and with 9 of these having $f_c > 0.95$.
The covering fraction distribution in knot B is less flat, but   has $f_c >0.9$ in 7 consecutive bins for $v\le-50$km/s, with the first 3 having 
$f_c > 0.95$. The highest covering fractions in knot C is seen at   $v \sim -90$\,km/s  with two bins having  $f_c  \gtrsim 0.6$. 

The \sif\ absorption has approximately the same shape as the \sit\ profile, and the covering fraction distribution mimics that of 
\sit , demonstrating that the outflowing medium is multiphase, consisting of both neutral and ionized gas. However, for all knots 
there is a \sif\ low level absorption  tail extending to larger blueshifted velocities than seen in \sit , although the flat shape of the 
spectrum as it blends into the continuum noise makes an exact measurement of the maximum velocity difficult. The \sif \ doublet
is however also known to be strong in early type (O and B) stars, where high velocity 
highly ionized winds are common revealing  blueshifted  \sif\ absorption extending from $\sim -100$ to $\lesssim -1000$ km/s 
\citep{grady1987,howarth1989,shepard2020}.  All three knots show P Cygni profiles in C\,{\sc iv} and N\,{\sc v},  and though not obvious 
could be there for \sif\ as well.
Hence the \sif\ absorption profiles towards the knots in \haro\ may plausibly also
have contributions from circumstellar absorption and  interpreting differences with respect to \sit\  should have this caveat in mind.
 
There are  some subtle details worthy of mention here: \\
	-- In knot B the Si{\sc ii} and Si{\sc iv} profiles profiles follow each 
other closely, but there is significant absorption in both profiles at redshifted (compared to \ha) velocities up to +150 km/s. 
This can be understood in terms of bulk motions in the multiphase ISM; the \ha\ velocity dispersion 
(Table \ref{table1}) corresponds to a FWHM of 210 km/s so blueshifted components of this magnitude could be foreseen, even 
when accounting for the fact that \ha\ also samples gas on the far side of knot B. Another possibility is that parts of the clumpy 
medium are falling in to knot B. The \sif\ absorption extends to blueshifted velocities of at least --700 km/s.  

 	-- In knot A we see no neutral gas at redshifted velocities, but a
 redshifted Si{\sc iv} component up to $v\lesssim 100$\,km/s (again consistent with the \ha\ velocity dispersion), which on the blue side  
 extends to $v \lesssim -750$\,km/s. 
 The low covering of neutral gas at zero velocity ($f_c\approx 0.3$) leads to a narrow window, on the blue side of the Lyman limit, 
 where LyC photons will not be absorbed.
 However this window is just 30 km/s, corresponding to 0.1 \AA, and hence insignificant for the escape of LyC. 
 
 	-- In knot C we see redshifted \sit\ absorption up to $v\sim +100$\,km/s,  but the \sif\ absorption extends to $+250$, and
--800\,km/s, on the red and blue sides, respectively. 
	
 The inferred neutral ISM covering is significantly smaller in knot C  than in A and B, but the ionized gas covering, while still smaller (the \sif\
absorption never goes to zero), is not markedly
different. Hence, while a similar amount of gas may be moving towards knot C as in knot B, the ionization/pressure balance is such 
that the once neutral clouds have largely evaporated.
The tail of blueshifted \sif\ gas with $v<-500$ km/s is seen in all knots, but is not detected in   \sit\ , suggesting that  neutral clouds at those 
velocities have been ionized, but could also be due to circumstellar  absorption from hot stars with fast winds.

In Fig. \ref{fig:grid}  we compare the column densities, covering fractions and \sit\ 1260 line profiles of knots A, B and C.
In the left panel, the reference velocity is the systemic  \ha\ velocity  for each individual knot (6251, 6170, 6140 km/s for knots A, B and C, respectively)
In the right panel, the zero point of the velocity scale is 6140 km/s, the systemic \ha\ velocity of knot C  (we will return to this  in Sect 4.5).
The derived column densities of Si\,{\sc ii} at $v_{min}$ (left panels of Fig. \ref{fig:grid}) is  $\log(N_{\rm SiII})=13.0$, 12.8, 12.4 (cm$^{-2} / $(km/s)) 
for knots A, B and C, respectively, 
or $\log(N_{\rm SiII})=14.2$, 14.0, 13.6 (cm$^{-2}$) when integrated over a 17 km/s velocity bin.  
However, the AOD method saturates at high optical depth, with the consequence that for several velocity bins in knots A and B, the upper limit on the 
column density is unconstrained,  and there may also be smaller high column density clouds immersed in the
moderately thick medium here probed. 
Following \citet{rivera-thorsen2017} we adopt a Si/H abundance ratio of $3.2\cdot10^{-6}$ implying that the neutral H column at minimum \sit\ intensity 
is of the order of $\log(N_{\rm HI})\sim 19.7$,  19.5, and 19.1 (cm$^{-2}$), for knots A, B and C, respectively. 
 The Lyman continuum becomes optically thick ($\tau_{{\scriptscriptstyle LyC}}=1$) for
$\log(N_{H})=17.2$, and the covered fractions of these velocity bins are completely opaque to both \lya\ and LyC.  
The covering fractions at $v_{min}$ are $f_c=95$, 96 and 55\%, and the LyC transmissions through the $v_{min}$ 
velocity bin hence 5, 4 and 45 \% for knots A, B and C, respectively.

We can in principle follow the same procedure for each velocity bin. However, since we do not know how the covered versus non-covered regions 
correlate spatially between velocity bins, we are unable to estimate the LyC transmission integrated along each sight line. If there is no correlation, 
the integrated transmission approaches zero after just a small number of  velocity bins.

We can estimate the number of absorbing clouds per line of sight required to produce the observed 
absorption and $f_c$ profiles. We first note that with expected cloud sizes on the order of $\lesssim1$ pc \citep{mccourt2018}, and effective
aperture diameters of $\sim 0.3\arcsec$ (area $\sim10^4$ pc$^2$) each aperture would contain $\gtrsim 
10^4$ independent unresolved lines of sight that could hit a cloud.  
If one assumes optically thick absorbers and a reasonable transonic
velocity dispersion inside the cold $\sim 10^4\,$K  medium in which
\sit\ is embedded\footnote{Note that while this is a fair assumption
for the ISM, absorption line studies of the CGM find routinely much
smaller Doppler parameters consistent with purely thermal broadening
\citep[e.g.][]{churchill2020}.}, one can estimate the expectation value for
the number of absorbers for a sightline towards the knots   as 

$$N_{\rm abs.}\approx \sum_i f_{\rm c,i} \Delta v/w_{\rm abs}$$

where $\Delta v$ is the spectral sampling (17 km/s), and  $w_{\rm abs}$ the
velocity width of an optically thick cloud. For $w_{\rm abs}$ we take the 
absorption equivalent width $w$ in velocity units which in the optically thick
regime is $w= 2 b \sqrt{\ln \tau_0}$ \citep{petitjean1998}, where $b$ is the doppler width (13 km/s for $10^4\,$K)
and $\tau_0$ the optical depth at line center. To estimate $\tau_0$ we take 
the average \sit\ column density ($\log N=12.7$ cm$^{-2}$/km s$^{-1}$)  times the 
spectral resolution ($\sim 2\times \Delta v$) and the oscillator strength for 
\sit$_{1260}$ to get $\tau_0\sim30$ and $w_{\rm abs}\sim50$ km/s ($\sim$0.2 \AA). 
This
yields $N_{\rm abs.}~\{8,\,8,\,4\}$ for clump A,B, and C,
respectively. Assuming furthermore an uncorrelated (Poissonian)
distribution of the absorbers in space, the probability of not
encountering one is simply $e^{-N_{\rm abs}}$ which yields
$\{< 0.1\%,\, <0.1\%, \sim2\%\}$ free lines of sight for the three clumps,
which can be regarded as lower limits for the escape fractions.

If, on the other hand, the non covered part of the aperture
is spatially correlated between velocity channels, there can be a significant net LyC transmission. Spatially correlated 'holes'
would be expected if these are produced by either mechanical feedback (winds) or carved out by ionizing photons, so this is more likely
in the current context, but the correlation is unlikely to be perfect. In the extreme case of perfect correlation, the escape fraction would be 
equal to that of the most covered velocity bin ($f_c =0.98^{+0.02}_{-0.05}, 1.0^{+0}_{-0.15}, 0.65^{+0.02}_{-0.07}$), hence escape fractions of $2^{+5}_{-2}$, $0^{+15}_{-0}$ and $35^{+7}_{-2}$\%, 
respectively for knot A, B and C.

With this caution in mind, and noting that some channels have unbounded upper limits on the column densities,  we can make a rough estimate for the effective total \hi\ columns contained in the solid angle towards each knot
by   taking 

$$N = \sum_i   N_i \times f_{c,i} \times \Delta v$$

 i.e the sum of the column densities times the covering fraction over the range of  velocities with reliable values (range [--500:0],  
[--440:150] and  [--430:120]  km/s for knots A, B and C, respectively; 
interpolating over  unreliable bins, such as +10 for knot A, and --220 for knot C).  
The results\footnote{The results differ from those of \citet{rivera-thorsen2017} where the integration over the width of each velocity 
bin, amounting to a factor of 17  corresponding to the width in km/s per bin, was omitted.} 
are  presented in Table \ref{tab:aod}. 	The integrated $f_c\times N$ of knot A and B corresponds to $\log(f_c\times N_{H})=20.7$
 (in rough agreement with the estimates from the Voigt profile fitting) and 20.2  log(cm$^{-2}$) for knot C.
 The estimate for knot C could be on the low side by $\sim0.3$ dex due to the lower O/H;  there is, however, is no sign of broad \hi\ 
 absorption near \lya , and we do not know whether the lower O/H applies over the full effective line-of-sight extent ($\sim 10$ kpc)
 or is local to knot C.

We also summed  $N_{i,\rm Si{\scriptscriptstyle IV}} \times f_{c,i}$ over the observed velocity width to arrive at similar, but somewhat lower, values as for \sit . 
However, this slightly underestimates the column density,  since the ionised medium  also contain some Si\,{\sc iii} and a little  Si\,{\sc v}.  
With a residual neutral hydrogen fraction in the ionized outflow component of $< 10^{-5}$ \citep{gray2019}, the integrated paths through 
the \sif\ gas would be optically thick for \lya ; hence it would scatter also in this medium, removing \lya\ from line center \citep[cf][]{verhamme2015}, 
but thin ($\tau <0.1$) to ionizing photons.

In addition to the \sit\ and \sif\ transitions used for the AOD analysis, we detect clear LIS absorption in \oi\ 1302 and \cii\ 1334+1335, and for 
knots A and B also S\,{\sc ii} 1250+1253. The \oi\ line is of paticular interest as the ionization potential of \hi\ and \oi\ are both 13.6 eV, and the matter probed
must come from the neutral hydrogen phase. For knots A and B, the \oi\ line has the same shape, width and depth (going to $\sim$ 0 intensity)
as the neighboring \sit\ 1304 line (see Fig. \ref{fig:lis}), while for knot C the width of \oi\ is similar to \sit\ although the absorption is much shallower ($\sim$0.8 times the local 
continuum for \oi\ compared to $\sim$0.6 for \sit).
 This lends support to the 
 conclusion that the covering fraction
towards knot C is significantly smaller than for knots A and B.

The \oi\ 1302 line has a oscillator strength of 0.049 compared to 0.086 for the neighbouring \sit\ 1304 transition. However, oxygen is
significantly more abundant  than silicon (typically log(Si/O)=-1.7, Garnett et al. 1995) so one would  expect  the oxygen line to 
be stronger (as seen in knot B), unless both are totally optically thick in which case they should be equally strong (as seen in knot A). The 
fact that the \oi\ line is weaker in knot C might indicate that \sit\ here partly arise in ionised gas and hence overestimates the neutral gas 
column and covering fraction.  Since the ionisation (\siii /\sii) is much lower in knot C a significant fraction of \sii \ and \sit\ may come 
from a phase where hydrogen is ionized. Thus, the neutral hydrogen covering fraction in knot C could be closer to 20\% .

\section{Discussion}

\subsection{Ionization structure and results by Keenan et al.}

\citet{keenan2017} used HST narrowband images to study the spatial pattern of ionization through 
 \oratio , \oii /\ha\ and \oiii /\ha . At the position of knot A,  \oii /\ha\ is low and \oiii /\ha\ high, suggesting a low
optical depth along the line of sight, and they argue that knot A may therefore be the source of LyC leakage from \haro . 

We deem it interesting here to  compare to the velocity sliced \oiii /\ha\ mapping based on MUSE data and presented 
in    \citet[][]{menacho2019}. While the ionization is indeed very high around knot
A, many of these structures  seem to extend in the plane towards S and W. 
Very high ionization  (\oiii /\ha $\gtrsim 3 $)  is prominent $>1$\,kpc SW of knot A  from --300 to --150 km/s  (and also at +250 km/s) \citep[][Fig. 3]{menacho2019} 
but the  ionization at the actual center of knot A
is moderate (\oiii /\ha $\gtrsim 1 $). Hence this structure  is likely outflowing, created by ionizing photons from knot A and optically thin to LyC photons, 
but pointing towards us at some angle,  not directly into the line of sight. 

However, between
--100 an 0 km/s there is a point like source of high ionization in  \citet{menacho2019}, $\sim 0.5$\,kpc W of the center of knot A (hence within the knot) and 
it reappears as highly ionized at $v>250$ to 350 km/s. It can be identified with the westernmost star cluster of knot A (see Fig. \ref{fig:acq} where it is indicated by a blue circle), and in Fig. \ref{fig:fig1} as the bright red (indicating strong \ha\ emission) source near the western edge of the COS aperture. The source is also clearly seen in the work of  \citet{keenan2017}:  in Fig.1 as the bright rightmost source in the box labelled knot A  and in the central panel of Fig. 5  as the source in knot A with \oiii / \oii $\sim 10$. 
	This source can also be identified with one of the young massive star clusters  (cluster \#15) in \citet{adamo2010} and has an estimated age of 3.5 Myr and mass of $5\cdot 10^{5}$\msun.  The  \ha\ velocity at this location is
$\sim +80$ km/s \citep[compared to the systemic velocity  6194 km/s of][coincident with the dusty arm]{menacho2021}, while the average for knot A is 57 km/s. 
In the kinematic study of \citet{ostlin2015} it was shown that the stellar and ionized 
gas velocities broadly agree in knot A, but also that the ionised gas here has several distinct velocity components due the projection onto the
more redshifted dusty arm \citep[][referred to as the 'ear' in \citet{ostlin2015}]{menacho2021}. Hence it is probable
that this cluster actually lies behind knot A, in the dusty arm.  However,  it also coincides with a local minimum in the  nebular reddening 
($E(B-V)_n\approx 0.16$), 
slightly lower than the  rest of knot A and the dusty arm.

The ionization of this source is moderate at its systemic velocity but increasing for more redshifted velocities. 
The ionization drops 
at  yet higher velocities which may due to blending with the overall halo gas that has somewhat lower ionization due to 
exhaustion of more energetic photons ($>35$ eV) capable of doubly ionizing oxygen.  

Cluster \#15 is the strongest \ha\ source within the knot A aperture, and accounts for approximately 30\% of  the \ha\ flux 
in the COS aperture, but only contributes $\sim$10\% of the far UV flux.    This means that the the UV continuum and \ha \
will have slightly different luminosity weighted mean velocities, and, as a consequence, we might expect a deficiency  of  ISM
absorption for velocities $v> -20$ km/s, which may affect the \sit\ absorption in knot A close to 0  km/s.

  Another 
high ionization structure, in the form of a cone,  is seen extending north from knot B at --50 to +100 with an opening angle of $\sim 5\degr$ and spatially extending to the boundary
of the galaxy, where the ionization even increases. This is most likely another LyC escape channel  directed roughly perpendicular
to the line of sight. \citep[][Fig. 4]{menacho2019}

Knot C shows low to moderate ionization at all velocities. 
Knot C is spatially resolved, hence not a bona fide star cluster, but more like a very
UV bright galaxy core or nuclear star cluster. Presumably it has experienced an extended star formation history, perhaps  for the past 40 Myr 
when the massive star cluster formation was triggered \citep{adamo2010}. 
Despite this, the electron temperature is highest  in knot C \citep[see further discussion on this  in ][]{menacho2021} 

\subsection{Dust geometry}
We return now to ${\mathcal R}$  the 'extinction corrected' \lya \ escape fractions.
In general, galaxies are observed to have 		
${\mathcal R}$ significantly lower than 1
\citep{giavalisco1996,ostlin2009,hayes2011,atek2014}, indicating that
part of \lya\ has  been scattered out of the line of sight and/or that scattering has increased the effective path length and hence the 
effective dust column. In the rare case  ${\mathcal R}>1$, then either {\bf i)} \lya\ has been  scattered into the line of sight, or {\bf ii)} 
has experienced a lower dust column or {\bf iii)}  the dust geometry is not in the form of a screen, but mixed with the sources,  
or in discrete clumps \citep{scarlata2009}. 
Case {\bf i}) is not  likely for spectra centered on bright UV continuum sources, such as knots A, B and C, since the \lya\ production outside
the apertures is much lower than within them,  but is a reason why   \lya /\ha \
$ >8.4$ is observed in the halo of many galaxies including \haro\  \citep{hayes2007,ostlin2009,ostlin2014}. Case ({\bf ii}) can happen due to geometrical
effects in the ISM, e.g. if dust is mainly contained in  dense clouds and \lya\ scatters on their surfaces but do not traverse them 
\citep{neufeld1991,laursen2013,duval2014} or similarly if there are ionized dust-poor channels where \lya\ scatters on the interfaces. 
For case {\bf iii})  the wavelength dependence on the optical depth will be softened \citep{scarlata2009}, effectively increasing \lya /\ha . It should be noted 
that cases {\bf ii}) and {\bf iii}) are not distinct: the dust is likely, and observed to be, inhomogeneously distributed but in a more complex fashion than 
having clouds of fixed opacity with no dust around them, and \lya\  will always have more complicated radiative transfer since it is be prone to scattering 
on \hi .

 From Table \ref{tab:results} we see that knot A has ${\mathcal R}\ll1$ and
hence \lya\ has been significantly attenuated by scattering increasing the effective dust optical depth. 
For knot C,  ${\mathcal R} >1$ and  \lya\ must have been scattered through less dusty paths in a clumpy 
two-phase medium or through ionized channels.   
Those paths with low \hi\ density would also be optically thin to LyC photons. 

Interestingly, this scenario is also consistent with less neutral
absorbers per line-of-sight for clump C as estimated in Sect \ref{AOD}. In
such a case \textbf{(ii)} described above is more likely as then the
majority of \lya\ photons will escape via `random walk' between the
clumps and, thus, are less susceptible to dust embedded within them.
If, on the other hand, the number of absorbers increases, more \lya\
photons will escape via `frequency excursion' and cross the absorbing clouds and their
dust \citep{gronke2017}.

\subsection{Stellar vs nebular reddening}  
The amount of reddening of the stellar component for young populations can be estimated from the 
UV continuum slope, $\beta$ ($f_\lambda \propto \lambda^\beta$) \citep{calzetti2000}.  To measure $\beta$ 
we used the derivation from \citet{micheva2020}  based on images in the  ACS/SBC/F140LP,  STIS/NUV/F25QTZ and WFC3/UVIS/WFC3 
filters and retrieved the value for each knot by  integrating over the COS aperture after having applied the COS vignetting function \citep[][]{goudfrooij2010}, 
and correcting for foreground Milky Way extinction (as above), to obtain
a $\beta$ representative for the observed COS spectrum of each knot. (see Table \ref{tab:results}).

A  value $\beta= -1.73$ indicates 0.85 magnitudes of extinction at 1600\AA , and $E(B-V)_n\approx0.19$ \citep[using equations 9 and 10 in][]{calzetti2000}. 
Hence, for knot A, the values of $E(B-V)_n=0.18$ and $\beta= -1.73$ are perfectly consistent with the \citet{calzetti2000} prescription that $E(B-V)_s = 0.44E(B-V)_n$.
For knot C the reddening implied from $\beta$ ($A_{1600}\lesssim0.5$) is a factor of 3.2 lower at 1600\AA\ than what $E(B-V)_n$
suggests ($A_{1600}\approx1.6$). 
For knot B, the situation is the reverse, with $\beta$  implying  2.9 magnitudes of extinction at 1600\AA\ while \ha /\hb\ 
suggests $A_{1600}\approx 1.7$ mag. 
The dust 
geometries are quite complex in knots B and C which is apparent from just looking at the HST multicolor image \citep{adamo2010},
where dust lanes streak though the centers of knots B and C. Possibly, in knot C, the dusty nebular gas emitting \ha\ and \hb\ only
covers a small part of the UV radiation (consistent with the low neutral gas covering fraction), while in knot B dust could preferentially 
be located in the UV luminous parts, surrounded by a 
bright emission line nebula with lower dust content? 
	Indeed, the \ha/\hb\ and Br\,$\gamma$/\hb\ measurements for knot B and C from XSHOOTER  in a 0.9\arcsec\ wide slit \citet{guseva2012}
differs significantly from our values, with the centre of knot B having a higher extinction: $E(B-V)_n = 0.51$ and 0.60 from \ha/\hb\ and 
Br\,$\gamma$/\hb\, respectively (implying $A_{1600}=2.3$, and 2.6, respectively), closer to the  $\beta$ estimate. 
Knot C instead
has a lower inferred $E(B-V)_n = 0.2$ (implying $A_{1600}=0.9$). 
These differences corroborates the conclusion that the dust distribution 
in knots B and C are highly non-uniform \citep[see also][]{menacho2021}.

The above discussion used the \citet{calzetti2000} law. If adopting instead the SMC law of \citet{prevot1984}, lower stellar extinctions
(about half) would result, mainly because the selective extinction ($R_V$) is lower, and in this case the far UV extinctions inferred from
$\beta$ and \ha/\hb\ in knot B would be largely consistent, while for knot  C the discrepancy would increase.

Anyway, an inhomogeneous dust distribution seems a like a natural explanation for both the small
stellar reddening and  high ${\mathcal R}$ of knot C, and lends support to 
knot C as a probable  source of Lyman continuum leakage. In knot B, geometry may have worked
in the opposite direction to preferentially redden the UV continuum, and the fairly high 
${\mathcal R}$ indicates that the average escaping \lya\ photon has encountered less dust 
than the UV continuum. 
We finally note that the individual clusters in knot A (including \#15) have bluer spectra ($\beta \lesssim -2$) than knot A integrated. 
The same applies for knot C where the centre has $\beta \approx -2.1$ \citep{micheva2020}.

\subsection{Ionizing photon budget} 
We now pose the question if any of the knots  have the required UV flux to be a potential source of the  LyC flux density observed with FUSE, which is $f_{\lambda}^{LyC}=4\pm0.9\cdot10^{-15}$ \egsa\ \citep{leitet2011}? 

The magnitude of the Lyman discontinuity depends on age and metallicity of the stellar populations, which we are not addressing in detail in the current paper. 
We note, however, that knot B is the youngest  with most clusters younger than 5 Myr, while knot A has a spread in age from a few to 10 Myr \citep{adamo2010}, 
and knot C likely has had an extended SFH for the past 40 Myr. Adopting a Starburst99 \citep{sb99} model with a continuous SFR and metallicity Z=0.008 (the 
results are very similar for Z=0.004) with age 10, 5 and 50 Myrs  leads to  Lyman discontinuities of $f_{\lambda912}^+ / f_{\lambda912}^- = 2.9, 2.3$ and 3.8 for 
knots A, B and C, respectively. Here, $f_{\lambda912}^+ / f_{\lambda912}^-$ is actually defined as $f_{\lambda1080-1120} / f_{\lambda870-900}$  \citep{sb99}.

We use the observed continuum flux densities at 1265\AA\ and the  $\beta$ values above to extrapolate to the expected flux densities at 1100\AA, and find $f_{\lambda912}^+ = 2.2\cdot 10^{-14}, \ 5.2\cdot 10^{-15}, {\rm and} \ 5.2\cdot 10^{-14}$ \egsa , for knots A, B and C, respectively.
We can then make the rough estimates that knots A, B and C would need to
have  escape fractions of $\sim$53\%,  $\sim$177\% and $\sim$29\%, respectively, to account for the measured LyC flux density. Hence knot B seems quite unlikely as a major contributor to the observed LyC, and also knot A in regard of the high neutral ISM covering. Knot C requires the least intrinsic escape and in addition has the lowest inferred ISM covering (with an estimated LyC transmission of 
2 to 35\%, see Sect. \ref{AOD}), making it plausible as a significant source of the escaping LyC. 

Zooming in on the  aforementioned cluster \#15 (the westernmost cluster in knot A), could it account for a  fraction of the observed LyC flux?
For an age of 3.5 Myr,  and a Starburst99 standard model \citep{sb99} with instantaneous burst and $Z=0.004$, $f_{\lambda912}^+ / f_{\lambda912}^- =2.9$,  and
even if 100\% of its LyC photons would escape it could not account for more than 25\% of the observed LyC flux. 
Since this source has $\sim 10\%$ of the far UV flux of knot A, a 100\% local escape would require a neutral covering fraction in knot A of
$f_c \lesssim 0.9$ while we derive $f_c > 0.95$ over a wide velocity range (150 km/s). 
Moreover, the \ha\ emission from this cluster is quite round and symmetric with a diameter of $\sim 0.5\arcsec$ (as judged from the HST \ha\ 
image) suggesting a quite high covering of the nebular gas. 
The equivalent width of \ha\ at cluster \#15 is indeed very high:
$EW({H\alpha})=1490$\AA\, integrated over the 0.5\arcsec\ diameter.  This value is close to what  
Starburst99 predicts (1320 \AA ) for the age in question, and hence most LyC photons seem to result in ionization implying \fesclyc $\ll 1 $.  
We conclude that cluster \#15 is not likely to contribute to the observed LyC flux.  
The eastern part of knot A has a more much more patchy and fractured \ha\ emission, and while $EW(H\alpha )$ is far from uniform, it is
on average lower (450 \AA ).  
It does in this respect seem more likely that the other parts (than cluster \#15) of knot A are leaking LyC photons to the SW parts of the
galaxy and powering the high ionization seen there \citep{keenan2017, menacho2019}.

While knot B (based on the \ha\ flux) produces more LyC photons than knots A and C together, it appears as quite unlikely as a major source of the detected LyC emission, since it would require the LyC photons to have some exclusive escape paths not available to the  UV continuum at $\lambda\sim1200$\AA . Neither the ISM covering, nor the extinction measurements suggest this to be the case. 
Hence, while there
may also be other sources to the observed LyC flux (e.g. recombinations directly to the ground state from the ionized halo, \"Ostlin et al. 2021, in prep.), and in principle more than one knot could be leaking, at least we can rule out knot B as a significant source of the observed Lyman continuum in \haro .  Knots A and C are both plausible sources, but the former requires a higher intrinsic escape fraction at odds with the ISM covering.

\subsection{Implications from the \lya\ profiles and AOD analysis}
The connection between \lya\ and LyC escape has been studied theoretically by \citet{verhamme2015} and empirically by
 \citet[][]{verhamme2017}.
The former work highlighted the presence of \lya\ emission at line center as a smoking gun for LyC escape. 
We see non-zero flux at $(1+v_{sys}/c)\times \lambda_{Ly\alpha,0}$ for all knots, especially for knot A and B, where the read peak is at $\sim+150$ km/s
and hence consistent with instrumental broadening of FWHM 150 km/s, implying that the detection of \lya\ at systemic velocity is not significant.
Moreover, the clumpy model employed by \citet{verhamme2015}
had zero neutral hydrogen density in the interclump medium, which is not realistic in view of ionised nebulae having neutral hydrogen fractions of $10^{-5}$ to
$\gtrsim10^{-4}$ \citep{agn2,gray2019} enough to make \lya\ optically thick. In  \citet{verhamme2017} and  \citet{izotov2018a,izotov2018b}  
the \lya\ properties of  confirmed LyC leakers were studied finding a positive correlation between \lya\ and LyC escape, and a negative correlation 
between  LyC escape and \lya\ peak separation. Peak separations of $\sim 400$ km/s, akin of knot B (and possibly C), are found in some confirmed LyC leakers with escape fractions $\lesssim10\%$ \citep{izotov2018b}. 
Based on the  \lya\ escape fractions, line widths and the peak separation,  it appears from these empirical relations that knot A is an unlikely source of LyC.
On the other hand, the radiative transport of \lya\ and LyC is very different: \lya\ bounces on the neutral clouds, but is still optically thick in the ionised
medium, and  \citet{jaskot2019} find no correlation between \fesclya\ and \oiii /\oii. However they find that the \lya\ peak separations correlate with \oiii /\oii \
and that \fesclya\ correlate with the LIS gas covering fraction, and suggest the LIS lines trace the fraction of low column density channels, and that
the ionization level is related to the transparency of those. In this context, this could mean that knot C has many free channels, but that they are relatively opaque. 
On the other hand the inferred \sif\ columns are not higher in knot C, which also has markedly lower (at least locally) electron densities 
\citep{james2013, menacho2021}.
The ISM absorption lines exclusively measure the gas properties along the line of sight  (which is what matters for LyC escape), 
and the ionization level has limited power in deducing the line of sight LyC optical depth since it is not affected by \hi , and may therefore have greater
diagnostic power on the sample level than for individual galaxies/star forming regions. 

\citet{kakiichi2019} also study the relation between LyC escape and \lya\ peak separation based on radiation hydrodynamics simulations, and in 
addition also the \lya\ red peak asymmetry (see their Fig 13, and Table \ref{tab:results}).
The roperties of knot A and B are consistent with those expected for the radiation bounded regime and low LyC escape. 
Knot C has a markedly higher red peak asymmetry which is consistent with the regime where LyC escapes through holes. 

\citet{chisholm2016} finds that metal absorption lines of similar strength but different ionization state (among them \sit \,{\small 1260} and \sif \,{\small 1393} used here) 
show similar velocity and line width suggesting they trace the same structures  \citep[see also][]{henry2015}. We arrive at a similar conclusion, although we see a 
somewhat wider
velocity distribution for the ionized gas than for the neutral component. 
 \citet{chisholm2018} 
showed that the \sit\ covering fractions are indeed
correlated with the \hi\ covering derived from the Lyman series (excluding \lya )  absorption and that this correlates with the observed LyC escape fraction.
Using the relation (which has a considerable scatter) between \sit\ and \hi\  covering in \citet{chisholm2018}  we expect knots A and B to have \hi\ covering fractions of unity, and
for knot C, $f_{c,HI}\sim 0.8$, and thus a net LyC escape. 

While all these relations have considerable scatter,  knot C  stands out as the most likely source of LyC.

Returning to the shape of the absorption profiles and the choice of reference velocity in Fig. \ref{fig:grid}, where in the right panel all profiles are
compared relative to the \ha\ velocity of knot C.
Knot B and C differ only $\sim$30 km/s in systemic \ha\ velocity (and less considering stellar velocities), and changing to individual reference velocities have little effect on the comparison.
Knot A, however, differs with $\sim$110 km/s from knot C and in the comparison in the right panel of Fig. \ref{fig:grid}  the absorption profile of 
knot A is significantly blueshifted with respect to knots B and C. The similarity of the absorption profiles (and also $N_{\rm Si{II}}$ and $f_c$) in the 
right panel of Fig. \ref{fig:grid} (in particular for A and B, agreeing both in shape and magnitude) is quite striking and suggest that they may be
reflecting  the same global outflow structure, which 
suggests that the absorption is perhaps not 
dominated by matter in the knots themselves, but further out,  along the line of sight towards us; the projected distance between knots A and B 
is just 1.4 kpc.
The absorption profile of knot C is 
significantly shallower but cover the same range of velocity.  \haro\ is a merger \citep{ostlin2015}, and likely the
division is such that knots A and B originate in the same late type galaxy (with very active star formation, many star clusters and dust lanes)
while knot C is from the other, less gas rich,  galaxy participating in the merger.  This division is also supported by kinematics as the emission
line profile halfway between knot C and A are markedly double peaked with two components separated by $\sim100$ km/s \citep{ostlin2015}.  

The  ionized halo of \haro\ as seen in the MUSE \ha\  image is quite round
and extends to a radius of 13 kpc \citep{menacho2019}. If we assume that the radius along the line of sight is similar and that this also equals 
the extent of the neutral outflows, which is observed to be $\gtrsim 500$ km/s, we  find that, on average each 17 km/s velocity bin would
correspond  to $\sim400$ pc. The individual neutral clouds probably have a radius on the 
order of $\lesssim$1 pc \citep{mccourt2018}, in which case the area of the aperture (taken to have a diameter of 0.3\arcsec) is $\sim10^4$ times 
the cross-section an individual cloud. The probability of a photon hitting a single cloud is then $10^{-4}$, and for 30\,000 randomly distributed clouds a 
covering fraction of $f_c=0.95$ would be expected, while 7000 clouds would yield $f_c=0.5$ and 1000 clouds $f_c=0.1$. The resulting volume 
filling fractions of neutral clouds per velocity bin would then be $\sim0.005$ (for $f_c=0.95$). These  are of course very 
crude estimates: the velocity vs radius scaling is not likely to be linear, and the cloud size  is likely smaller at small radii,
but nevertheless illustrates that high covering fractions may be found for low volume filling fractions.

\section{Conclusions}
We have presented new ultra violet HST/COS spectroscopy of knots A and B in \haro , and investigated the \lya\ line profiles and absorption lines from silicon
arising in the neutral  (\sit) and ionized (\sif) ISM. We combine these with available HST/COS spectroscopy for knot C, to make a comparative analysis of which knot, as inferred from the neutral ISM covering, might  be the most likely source of the  Lyman continuum escape detected in this galaxy. We have included optical data from the ESO/VLT/MUSE integral field spectrometer to derive the nebular extinction (from \ha /\hb), ionization level and \lya\ escape fractions for knots A, B, and C. 

All three knots show prominent \lya\ line emission detected at high signal to noise. Knot A and B have double peaks in \lya\ while 
knot C only shows a hint of a blue peak (or more likely a P-Cyg profile). Broad H{\sc i} absorption near \lya\ is seen in knots A and B, but not in knot C.
The \sit\ and \sif\ lines for all knots reveal outflowing material with blue shifted velocities up to $\gtrsim 500$ km/s. 
 From the \sit\ and \sif\ lines, we have derived column densities and covering fractions for the intervening neutral and ionized gas, respectively, and as a function of velocity. Below we summarize our conclusions:

-- Knot A has the smallest nebular reddening, and a high ionization, but yet has  a \lya\ escape fraction of only $\sim 1$\% , significantly lower than its reddening suggests, indicating that the \lya\ has been strongly attenuated by dust through multiple resonant scatterings. The \lya\ line is double peaked, but
with a line width and peak separation that indicates a rather high \hi\ column density. Broad absorption around \lya , analyzed by Voigt profile fitting rendered an \hi\ column of log($N_{HI})=20.7$ (cm$^{-2}$). The ISM absorption lines reveal  covering fractions of neutral gas close to unity over $\gtrsim 300$ km/s. While the high ionization in and around knot A led \citet{keenan2017} to suggest it as the most likely source of leaking LyC radiation, the high neutral covering  imply that significant  LyC escape along the line of sight is unlikely.

-- Knot B has the highest observed \ha\ flux and equivalent width, but also the highest reddening, hence this is where most LyC and \lya\ photons are
produced, but the \lya\ escape fraction is  $<1$\%. The \lya\ line width is similar to knot A, but the peak separation is smaller, with values comparable 
to some confirmed LyC leaking galaxies, but the Voigt profile fitting returns an \hi\ column , log($N_{HI})=21$, higher than for knot A. Moreover, the ISM absorption lines reveal  a covering 
fraction of neutral gas very similar to knot A. Also the UV continuum slope ($\beta$) is quite red, indicating even higher reddening than estimated from \ha /\hb .
In conclusion, significant LyC escape along the line of sight from knot B is quite unlikely.

-- Knot C has the highest \lya\ flux, and has an order of magnitude higher \lya\ escape fraction than knot A and B, despite a high nebular reddening
(similar to that of knot B). 
Curiously, the UV continuum slope is very blue indicating a significantly lower reddening than the nebular estimate. In fact, a lot more \lya\ photons
escape knot C than the nebular reddening would suggest, indicating that \lya\ escapes through less dusty paths than \ha .  The \lya\ profile is narrow, 
 and with no sign of 
broad absorption, all suggesting low \hi\ column density. The \sit\ lines suggest a  low {$\sim 50$\%} covering fraction of neutral gas, suggesting
a significant transmission (2 -- 35\%) of Lyman continuum along the line of sight.  The \oi\ 1302 \AA\ line is however shallower than the \sit\ lines
suggesting that the former may be formed also in the phase where hydrogen is ionized, and that the neutral covering may be even smaller, around 20\% .
Based on the ionization budget, knot C would need to have a LyC escape fraction of 30\% to account for all the observed LyC flux detected in the 
FUSE observations, and our results suggest this is plausible.

In conclusion, the observations  here presented and  analyzed,  favors knot C as  the most likely of the three knots to significantly contribute  to the observed LyC leakage from \haro .

\acknowledgments
This research has made use of the NASA/IPAC Extragalactic Database (NED) which 
is operated by the Jet Propulsion Laboratory, California Institute of
Technology, under contract with the National Aeronautics and Space
Administration.  This work was  supported by the Swedish Research Council (Vetenskapsr\aa det).
and the Swedish National Space Agency (SNSA).  
M.S.O. acknowledges support from NASA, HST-GO-15352.002-A. 
 D.K. is funded by the Centre National d'\'{E}tudes Spatiales (CNES)/Centre National de la Recherche Scientifique (CNRS); Convention No 131425.  
MMH is funded by Spanish State Research Agency grant MDM-2017-0737 (Unidad de Excelencia Maria de Maeztu CAB).
MG was supported by NASA through the Hubble Fellowship grant HST-HF2-51409. 
The Cosmic Dawn Center is funded by the Danish National Research Foundation under grant No. 140.
We thank the anonymous referee for a thorough and constructive review.

\facility{HST (ACS,COS), ESO (VLT/MUSE)}

%
%


\bibliographystyle{aasjournal} 

\clearpage

\end{document}